\documentclass{aa}

\usepackage{graphicx}
\graphicspath{{figures}}
\usepackage{newtxtext, newtxmath}
\usepackage{placeins}           % useful with \FloatBarrier, to keep 
\usepackage{hyperref}
\hypersetup{breaklinks=true, colorlinks=true, allcolors=blue}
\usepackage{microtype}
\usepackage{orcidlink}
\usepackage{siunitx}
\DeclareSymbolFont{upgreek}{U}{eur}{m}{n}
\DeclareMathSymbol{\umu}{0}{upgreek}{"16}
\DeclareSIPrefix{\micro}{\text{\ensuremath{\umu}}}{-6}

\let\oldbibliography\thebibliography
\renewcommand{\thebibliography}[1]{%
    \oldbibliography{#1}%
    \setlength{\itemsep}{0pt}%
}

\begin{document}

\title{%
    Wavelength-dependent far-infrared polarization of HL~Tau\\ observed with SOFIA/HAWC+
}

% \subtitle{Subtitle}

\titlerunning{%
    Wavelength-dependent far-infrared polarization of HL~Tau
}

\author{
    Moritz Lietzow-Sinjen\thanks{Corresponding author; \href{mailto:mlietzow@astrophysik.uni-kiel.de}{\tt mlietzow@astrophysik.uni-kiel.de}}\,\orcidlink{0000-0001-9511-3371}
    \and
    Sebastian Wolf\,\orcidlink{0000-0001-7841-3452}
    \and
    Robert Brunngräber\,\orcidlink{0000-0002-3353-5277}
}

\authorrunning{%
    M. Lietzow-Sinjen~\etalname
}

\institute{
    Institute of Theoretical Physics and Astrophysics,
    Kiel University, Leibnizstr. 15, 24118 Kiel, Germany
}

\date{%
    Received 28 March 2024 / Accepted 21 January 2025
}

\abstract{
    We present the first polarimetric observations of a circumstellar disk in the far-infrared wavelength range.
    We report flux and linear polarization measurements of the young stellar object HL~Tau in the bands A (\SI{53}{\um}), C (\SI{89}{\um}), D (\SI{155}{\um}), and E (\SI{216}{\um}) with the High-resolution Airborne Wideband Camera-plus (HAWC+) on board of the Stratospheric Observatory for Infrared Astronomy (SOFIA).
    The orientation of the polarization vectors is strongly wavelength-dependent and can be attributed to different wavelength-dependent polarization mechanisms in the disk and its local environment.
    In bands A, C, and D (\SI{53}{\um} to \SI{155}{\um}), the orientation of the polarization is roughly consistent with a value of \ang{114} at the maximum emission.
    Hereby, the magnetic field direction is close to that of the spin axis of the disk.
    In contrast, in band E (\SI{216}{\um}), the orientation is nearly parallel to the minor axis of the projection of the inclined disk.
    Based on a viscous accretion disk model combined with a surrounding envelope, we performed polarized three-dimensional Monte Carlo radiative transfer simulations.
    In particular, we considered polarization due to emission and absorption by aligned dust grains, and polarization due to scattering of the thermal reemission (self-scattering).
    At wavelengths of \SI{53}{\um}, \SI{89}{\um}, and \SI{155}{\um}, we were able to reproduce the observed orientation of the polarization vectors.
    Here, the origin of polarization is consistent with polarized emission by aligned non-spherical dust grains.
    In contrast, at a wavelength of \SI{216}{\um}, the polarization pattern could not be fully matched, however, applying self-scattering and assuming dust grain radii up to \SI{35}{\um}, we were able to reproduce the flip in the orientation of polarization.
    We conclude that the polarization is caused by dichroic emission of aligned dust grains in the envelope, while at longer wavelengths, the envelope becomes transparent and the polarization is dominated by self-scattering in the disk.
}


\maketitle
\nolinenumbers

%%%%%%%%%%%%%%%%%%%%%%%%%%%%%%%%%%%%%%%%%%%%%%%%%%%%%%%%%%%%%%

\section{Introduction}
\label{sec:introduction}

\begin{table*}
    
    \centering
    \caption{General properties of the observations.}
    \label{tab:observation-properties}

    \renewcommand{\arraystretch}{1.2}
    \small
    \begin{tabular*}{\linewidth}{
        @{\extracolsep{\fill}}
        c
        S[table-format=3]
        S[table-format=2.1]
        S[table-format=2.2]
        S[table-format=1.2]
        S[table-format=5]
        c
    }
        \hline\hline
        HAWC+ band & {Band center} & {Band width\tablefootmark{a}} & {Beam FWHM} & {Pixel scale} & {Exposure time} & Date of observation \\
        & {(\si{\um})} & {(\si{\um})} & {(\si{arcsec})} & {(\si{arcsec})} & {(\si{s})} & (yyyy-mm-dd) \\
        \hline
        A &  53 &  8.7 &  4.84 & 1.21 &   960 & 2021-11-03 \\
        C &  89 & 17   &  7.8  & 1.95 &   694 & 2021-11-05 \\
        D & 155 & 34   & 13.6  & 3.40 &  3042 & 2021-09-08 \\
        E & 216 & 44   & 18.2  & 4.55 & 10525 & 2021-09-08 \\
        \hline
    \end{tabular*}
    \tablefoot{
        \tablefoottext{a}{Band width from \citet{Harper-etal-2018}.}
    }

\end{table*}

\object{HL Tau} is a pre-main-sequence star in the Taurus molecular cloud at a distance of about \SI{140}{pc} \citep{Kenyon-etal-1994, Rebull-etal-2004, Galli-etal-2018}, which is surrounded by a circumstellar disk with the possible formation of planets being observed \citep{Greaves-etal-2008}.
In addition, it is embedded in a circumstellar nebulosity that is optically thick at optical to mid-infrared wavelengths, making the object not directly visible in this wavelength region.
The visual extinction is estimated from \SI{22}{mag} to about \SI{30}{mag} \citep{Beckwith-Birk-1995, Stapelfeldt-etal-1995, Close-etal-1997}.
However, this system offers an opportunity to study the interplay between the still evolving circumstellar material and magnetic fields, which play an important role in star formation, as well as during the accretion process in the early stages and evolution of young stellar objects \citep[e.g.,][]{McKee-Ostriker-2007, Li-etal-2014, Tsukamoto-2016, Wurster-Li-2018, Hull-Zhang-2019, Pattle-etal-2023}.
Furthermore, jets and outflows were reported with a position angle consistent with the minor axis of the projection of the inclined circumstellar disk \citep{Mundt-Fried-1983, Mundt-etal-1990}.

At millimeter and submillimeter wavelengths, the object has been extensively studied \citep[e.g.,][]{Beckwith-etal-1990, Mundy-etal-1996, Looney-etal-2000, Kwon-etal-2011, Stephens-etal-2014, Lin-etal-2024}.
In particular, in observations with the Combined Array for Research in Millimeterwave Astronomy (CARMA) or the Plateau de Bure Interferometer (PdBI), properties such as the disk mass or the density distribution could be estimated \citep{Guilloteau-etal-2011, Kwon-etal-2011, Kwon-etal-2015}.
In addition, instruments such as the Atacama Large Millimeter/submillimeter Array (ALMA) provide additional properties of the circumstellar disk in recent years, such as ring structures or the distribution and alignment of dust grains \citep[e.g.,][]{ALMA-etal-2015, Pinte-etal-2016, Stephens-etal-2017, Carrasco-Gonzalez-etal-2019}.
The circumstellar disk itself is inclined by an angle of \ang{46.72 \pm 0.05} (from face-on) and has a position angle of \ang{138.02 \pm 0.07} east-of-north \citep{ALMA-etal-2015}.
The mass of the disk is estimated to be about \SI{0.1}{M_\odot} \citep{Guilloteau-etal-2011, Kwon-etal-2015}, \SI{0.13}{M_\odot} \citep{Kwon-etal-2011}, or even up to \SI{0.3}{M_\odot} \citep{Carrasco-Gonzalez-etal-2016}.

Moreover, polarimetric observations provide an opportunity to determine grain properties and to trace grain alignment in the disk and the circumstellar environment.
Hereby, polarization is expected to arise from two different mechanisms.
The first mechanism is self-scattering, that is, thermal radiation emitted and scattered by dust grains in the disk \citep[e.g.,][]{Kataoka-etal-2015, Brunngraber-Wolf-2019}.
The second mechanism is dichroic emission or absorption by non-spherical particles that align with a magnetic field, an ambient gas flow, or the radiation field \citep[e.g.,][]{Lazarian-2007, Andersson-etal-2015, Tazaki-etal-2017}.
Polarimetric millimeter and submillimeter observations with ALMA of several protoplanetary disks show polarization patterns that are consistent with the mechanism of self-scattered radiation, such as for IM~Lup \citep{Hull-etal-2018}, HD~163296 \citep{Dent-etal-2019, Ohashi-Kataoka-2019}, CW~Tau and DG~Tau \citep{Bacciotti-etal-2018}, whereas in some cases self-scattered radiation and radiation from aligned grains are proposed for the origin of polarization, such as for HD~142527 \citep{Kataoka-etal-2016b, Ohashi-etal-2018}, AS~209 \citep{Mori-etal-2019}, and HL~Tau \citep{Kataoka-etal-2017, Stephens-etal-2017, Mori-Kataoka-2021, Stephens-etal-2023}.
In particular, for the protoplanetary disk of HL~Tau, the polarization pattern at submillimeter wavelengths is consistent with a combination of self-scattering and alignment-induced polarization.
Furthermore, mid-infrared polarization due to dichroic emission of elongated particles is found in the disk of AB~Aur \citep{Li-etal-2016}.

In general, grain alignment is caused by torques acting on irregularly shaped dust grains.
Due to some external force such as an anisotropic radiation field (radiative torque; RAT) or a gas flow (mechanical torque; MET), dust grains start spinning and precessing around an axis given by either a magnetic field, the gas flow, or the radiation field \citep[e.g.,][]{Draine-Weingartner-1996, Cho-Lazarian-2007, Lazarian-2007, Lazarian-Hoang-2007a, Lazarian-Hoang-2007b, Hoang-Lazarian-2009, Andersson-etal-2015, Tazaki-etal-2017}.
In the case of RAT alignment, the rotating grains obtain a net internal magnetization due to the Barnett effect \citep{Barnett-1915} and finally align with their longer axis perpendicular to magnetic field lines while precessing around the field direction.
Thus, the orientation of polarization allows the underlying magnetic field morphology to be traced, for instance, in molecular clouds \citep{Reissl-etal-2017, Seifried-etal-2020, Zielinski-etal-2021, Zielinski-Wolf-2022}.
In contrast, if dust grains are aligned with an ambient gas flow, the longer axis can be either parallel \citep{Gold-1952} or perpendicular \citep{Lazarian-Hoang-2007b,Kataoka-etal-2019} to the drift velocity.

However, polarimetric measurements of protoplanetary disks in the far-infrared are not available so far.
In this study, we report the first polarimetric observations of HL~Tau in this wavelength range using the High-resolution Airborne Wideband Camera-plus \citep[HAWC+;][]{Harper-etal-2018} on board of the Stratospheric Observatory for Infrared Astronomy \citep[SOFIA;][]{Temi-etal-2018}.
We present polarization maps in the HAWC+ bands A (\SI{53}{\um}), C (\SI{89}{\um}), D (\SI{155}{\um}), and E (\SI{216}{\um}).
Furthermore, to analyze and discuss the polarimetric observations, we performed three-dimensional Monte Carlo polarized radiative transfer simulations.
Taking into account various polarization mechanisms, we aim to reproduce the observed degree and orientation of polarization at all four considered wavelength bands.

This paper is organized as follows.
In Sect.~\ref{sec:observations}, we outline the data acquisition and reduction.
Subsequently, we present and describe the polarization maps of HL~Tau in Sect.~\ref{sec:results}.
Afterward, we model the observational results using radiative transfer simulations in Sect.~\ref{sec:modeling}.
Finally, we summarize the study in Sect.~\ref{sec:conclusions}.

%%%%%%%%%%%%%%%%%%%%%%%%%%%%%%%%%%%%%%%%%%%%%%%%%%%%%%%%%%%%%%

\section{Observations}
\label{sec:observations}

As part of SOFIA Cycle 9 (ID: 09\_0084, PI: R.~Brunngräber), the observations were carried out using HAWC+ on flights 779, 786, and 788 on September 8, November 3, and November 5, 2021, respectively.
All observations were performed using the on-the-fly mapping (OTFMAP) polarimetric mode.
Table~\ref{tab:observation-properties} summarizes the general properties of the observations.

The raw data were reduced by the SOFIA Science Center staff using the HAWC+ data reduction pipeline (see \citealt{Harper-etal-2018} or HAWC+ DRP user's manual%
\footnote{%
    \url{https://irsa.ipac.caltech.edu/data/SOFIA/docs/sites/default/files/2022-12/hawc_users_revL.pdf}
}
for a detailed description of the pipeline and the data processing steps).
In this article, we use {level 4} data that are fully calibrated data, processed with version 3.2.0 of the HAWC+ pipeline.
The resulting data are FITS files that include Stokes $I$ (total intensity), $Q$, $U$, the degree of polarization $p$, debiased polarization degree $p'$, angle of polarization $\theta$, and the related measurement uncertainties $\sigma$.
The degree of polarization is given by
\begin{equation}
    p = \frac{\sqrt{Q^2 + U^2}}{I}.
\end{equation}
Following the approach of \citet{Wardle-Kronberg-1974}, the debiased degree of polarization is calculated as
\begin{equation}
    p' = \sqrt{p^2 - \sigma_p^2},
\end{equation}
where $\sigma_p$ is the error of the degree of polarization.
Finally, the polarization angle $\theta$ is defined by
\begin{equation}
    \tan(2\theta) = \frac{U}{Q}.
\end{equation}
For a detailed description, we refer to \citet{Gordon-etal-2018}.
We apply two quality thresholds for the following analysis, considering only data for polarization which satisfy the relations
\begin{equation}
    \label{eq:quality-cut}
    \frac{I}{\sigma_I} > 100, \quad \frac{p'}{\sigma_p} > 3.
\end{equation}

%%%%%%%%%%%%%%%%%%%%%%%%%%%%%%%%%%%%%%%%%%%%%%%%%%%%%%%%%%%%%%

\section{Results}
\label{sec:results}

\begin{figure*}

    \begin{center}

    \begin{minipage}{0.49\linewidth}
        \centering
        \includegraphics[width=\linewidth]{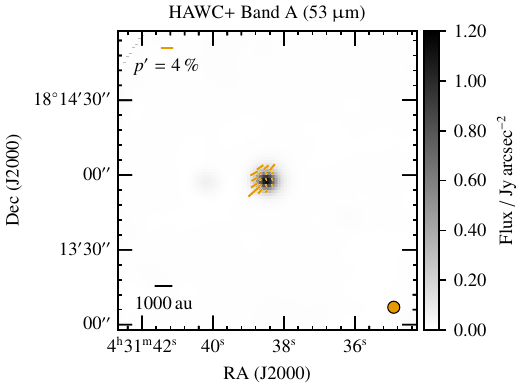}
    \end{minipage}
    \hfill
    \begin{minipage}{0.49\linewidth}
        \centering
        \includegraphics[width=\linewidth]{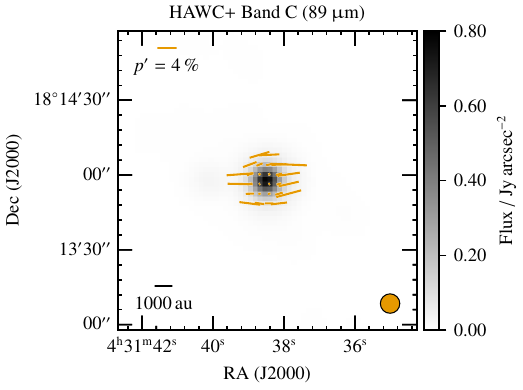}
    \end{minipage}

    \vspace*{0.8em}

    \begin{minipage}{0.49\linewidth}
        \centering
        \includegraphics[width=\linewidth]{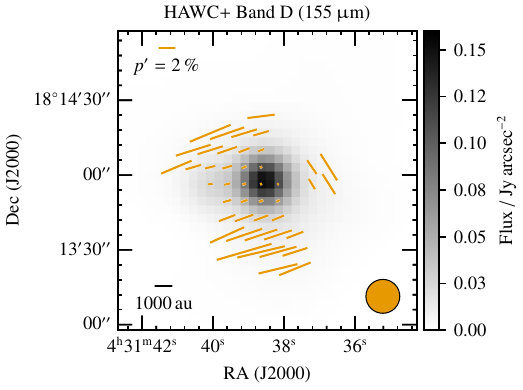}
    \end{minipage}
    \hfill
    \begin{minipage}{0.49\linewidth}
        \centering
        \includegraphics[width=\linewidth]{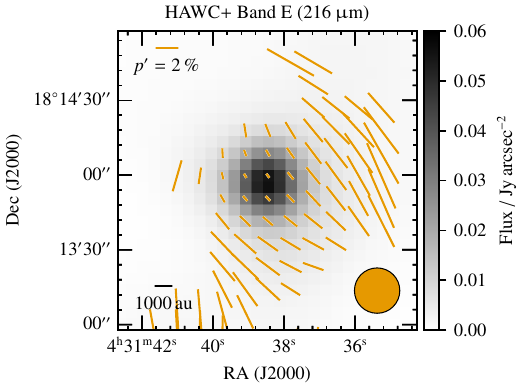}
    \end{minipage}

    \end{center}

    \caption{
        SOFIA/HAWC+ polarization maps of HL~Tau in band A (\SI{53}{\um}, {top left}), band C (\SI{89}{\um}, {top right}), band D (\SI{155}{\um}, {bottom left}), and band E (\SI{216}{\um}, {bottom right}).
        The surface brightness (gray scale) is given in units of \si{Jy.arcsec^{-2}} and is overlaid with polarization vectors in orange.
        The length and orientation of the vectors give the degree and angle of polarization, respectively.
        Only vectors that satisfy the selection criteria~\eqref{eq:quality-cut} are shown.
        The beam size (FWHM) at each corresponding SOFIA/HAWC+ wavelength band is indicated in the respective lower right corner.
        The scale bar (\SI{1000}{au}) corresponds to an assumed distance of HL~Tau of \SI{140}{pc}.
        The images are cropped to an image size of \SI{2}{arcmin} $\times$ \SI{2}{arcmin}.
        See Sect.~\ref{subsec:polarization-maps} for details.
    }
    \label{fig:sofia-polmap}

\end{figure*}

\subsection{Polarization maps}
\label{subsec:polarization-maps}

Figure~\ref{fig:sofia-polmap} shows the resulting polarization maps obtained in the four HAWC+ bands.
The images are centered at a position of about 04\textsuperscript{h}31\textsuperscript{m}38.5\textsuperscript{s}, $+$\ang{18;13;58} (J2000) and cropped to an image size of \SI{2}{arcmin} $\times$ \SI{2}{arcmin}.
The surface brightness is overlaid with polarization vectors that satisfy the criteria~\eqref{eq:quality-cut}.
In general, linear polarization of the system is detected in all bands and multiple polarization vectors fulfill the criteria.
Hereby, the length and orientation of the vectors indicate the degree and angle of polarization, respectively.
An angle of \ang{0} corresponds to the north-south direction with a positive sign in counterclockwise rotation.
The less luminous system \object{XZ Tau} is barely visible in the images at a position of about 04\textsuperscript{h}31\textsuperscript{m}40.1\textsuperscript{s}, $+$\ang{18;13;57} (J2000).

As listed in Table~\ref{tab:observation-properties}, with about \SI{1.21}{arcsec}, band A offers the smallest pixel scale.
However, at this resolution, the disk of HL~Tau is located roughly inside a single pixel.
Due to beam convolution, the intensity and polarization vectors cover an area on the image that is larger than the diameter of the circumstellar disk \citep[$\sim$\,\SI{1.5}{arcsec} at a wavelength of \SI{1.3}{mm};][]{ALMA-etal-2015}.
Consequently, the observed polarization at all four wavelength bands is likely a combination of the contribution of radiating dust grains in the disk itself and the large envelope surrounding HL~Tau.
However, with increasing wavelength, the optical depth of the envelope decreases, so the circumstellar disk becomes the dominant source of radiation.
In particular, in the submillimeter and millimeter wavelength regimes, the influence of the envelope is negligible \citep{Lay-etal-1997}.
We note that in bands D and E, the neighboring system XZ~Tau potentially also contributes to the polarization because of the relatively large beam size compared to the apparent separation of both systems.

In band A with a central wavelength of \SI{53}{\um}, the maximum emission amounts to \SI[allow-quantity-breaks]{1164.7 \pm 1.1}{mJy.arcsec^{-2}}.
The orientation of the polarization vectors in this inner region is about \ang{117 \pm 2} and the degree of polarization is in average \SI{0.49 \pm 0.04}{\percent}.
The highest degree of polarization is about \SI{5.1}{\percent}.

Furthermore, in band C with a central wavelength of \SI{89}{\um}, the intensity decreases and the maximum emission amounts to \SI[allow-quantity-breaks]{776.9 \pm 0.6}{mJy.arcsec^{-2}}.
This inner region shows an angle of polarization of about \ang{112 \pm 2}.
However, the polarization vectors in the outer regions mainly form an elliptical distribution.
In addition, the degree of polarization decreases with increasing total intensity.
The highest degree of polarization is \SI{7.5}{\percent} while it decreases to a mean value of \SI{0.40 \pm 0.02}{\percent} in the inner region.

Next, in band D with a central wavelength of \SI{155}{\um}, the maximum emission amounts to \SI[allow-quantity-breaks]{151.70 \pm 0.07}{mJy.arcsec^{-2}}.
The angle of polarization in this region is \ang{114 \pm 2}.
In addition, some vectors in the west of the image are flipped by about \ang{90} and are thus oriented almost along the minor axis of the projection of the inclined disk.
Similarly to band C, the degree of polarization decreases with increasing total intensity.
The highest degree of polarization is \SI{6.7}{\percent} while in the inner region, it is on average \SI{0.26 \pm 0.01}{\percent}.

Finally, in band E with a central wavelength of \SI{216}{\um}, the maximum emission amounts to \SI[allow-quantity-breaks]{56.17 \pm 0.03}{mJy.arcsec^{-2}}.
The orientation of the polarization vectors in the central region amounts to \ang{37 \pm 1}, which is almost consistent with the minor axis of the inclined disk projection.
The highest degree of polarization is about \SI{8.1}{\percent}.
In the inner bright region, the degree of polarization decreases to a mean value of about \SI{0.51 \pm 0.02}{\percent}.
In contrast to the bands at shorter wavelengths, there are also polarization vectors in the outer regions that satisfy criteria~\eqref{eq:quality-cut}.
The shape of this polarization pattern appears to be elliptical, similar to the situation in band C, but with a different orientation.

In conclusion, the maximum emission decreases with increasing wavelength.
In addition, the degree of polarization is smallest in the central region at maximum emission and increases with decreasing intensity.
Furthermore, the orientation of the polarization vectors is wavelength-dependent.

\subsection{Origin of polarization}
\label{subsec:origin-polarization}

While for bands A, C, and D the orientation of polarization at the maximum emission is about \ang{114}, it is about \ang{37} in band E.
The first orientation ($\sim$\,\ang{114}, bands A, C, and D) is close to the major axis of the inclined disk projection with a deviation of \ang{24}.
We conclude that the polarization is due to emission of aligned non-spherical grains.
However, if the grains are aligned with a magnetic field, the field direction has to be misaligned with the disk spin axis.
In contrast, the second orientation ($\sim$\,\ang{37}, band E) is almost parallel to the minor axis of the inclined disk projection with a deviation of only \ang{11}.
This orientation is consistent with the polarization due to self-scattering of an inclined disk \citep{Kataoka-etal-2016a, Yang-etal-2016, Brunngraber-Wolf-2019}.

\subsection{Mid-infrared and millimeter polarization}
\label{subsec:mir-mm-polarization}

In the mid-infrared wavelength region at \SI{8.7}{\um}, \SI{10.3}{\um}, and \SI{12.5}{\um}, measurements show that the polarization vectors are oriented with a position angle of about \ang{90} \citep{Li-etal-2018}.
The authors find absorption as a primary source of polarization while ruling out scattering.
Thus, polarization could potentially arise from aligned dust grains in the surrounding envelope.
This orientation would fit the elliptically shaped polarization vectors at the outer regions of our observations in band C.
However, the magnetic field direction would also have to be misaligned with the orientation and axis of the circumstellar disk.

At a wavelength of \SI{870}{\um}, the orientation of polarization is found to be parallel to the minor axis of the projection of the inclined disk \citep{Stephens-etal-2017}.
This polarization is expected to be due to scattering, and our observations at \SI{216}{\um} support this theory, since the polarization vectors are oriented parallel to the minor axis of the projection of the inclined disk.
At millimeter wavelengths, that is, up to a wavelength of \SI{7}{mm}, the polarization transforms to a pattern that is consistent with grains aligned toroidally \citep{Lin-etal-2024}.
In addition, high-resolution polarimetric observations at a wavelength of \SI{870}{\um} show an azimuthal component of the polarization angles in the gaps of the disk, suggesting the emission of aligned grains there as well \citep{Stephens-etal-2023}.
In the disk itself, the gaseous damping timescale dominates, thus other grain alignment mechanisms, such as alignment to the radiative flux \citep{Tazaki-etal-2017, Yang-etal-2019} or gas flow \citep{Kataoka-etal-2019}, are more likely.
For the circumstellar disk of HL~Tau, millimeter observations indicate alignment with the radiation field \citep{Kataoka-etal-2017, Stephens-etal-2017}.

%%%%%%%%%%%%%%%%%%%%%%%%%%%%%%%%%%%%%%%%%%%%%%%%%%%%%%%%%%%%%%

\section{Modeling the observations}
\label{sec:modeling}

In order to model the disk and to simulate the radiative transfer, we used the publicly available three-dimensional Monte Carlo radiative transfer code POLARIS%
\footnote{%
    \url{https://github.com/polaris-MCRT/POLARIS}
}
\citep{Reissl-etal-2016, Reissl-etal-2018}, which has been extensively tested and applied to a broad range of astrophysical models.
The code is capable of handling various polarization mechanisms such as dust scattering or absorption and thermal reemission of aligned grains \citep{Brunngraber-Wolf-2019,Brunngraber-Wolf-2020, Brunngraber-Wolf-2021}.
Throughout this study, we assumed a fixed angle of inclination of the disk from face-on of \ang{46.7} and a position angle of \ang{138} \citep{ALMA-etal-2015}.

Based on the results in Sect.~\ref{sec:results}, our main goal is to confirm the apparent underlying polarization mechanisms.
Thus, as a first step, we aim to reproduce the wavelength-dependent orientation of the polarization vectors.
For this purpose, we build a protoplanetary disk model, for which the polarization in bands A, C, and D is primarily caused by aligned dust grains, and the polarization in band E primarily by self-scattering (see Sect.~\ref{subsec:origin-polarization}).

We find that the polarization in each band can be reproduced separately by a single model by adjusting the mass of the envelope or the magnetic field properties, thus controlling the amount of polarization due to emission or self-scattering. 
However, it was difficult to construct a self-consistent model for the polarization at all bands.
In particular, a significant change in the orientation of the polarization from band D to E, that is, an increase in wavelength by only a factor of about \num{1.4}, is hardly achieved using a circumstellar disk alone.
Instead, we obtained a transition pattern of the polarization vectors as it is expected for such a small wavelength change.
Similarly, at a wavelength of \SI{1.3}{mm}, evidences for a polarization mechanism transition are found in ALMA observations \citep{Stephens-etal-2017}.

Furthermore, a large visual extinction of about \SI{22}{mag} to approximately \SI{30}{mag} is observed due to a surrounding envelope \citep{Beckwith-Birk-1995, Stapelfeldt-etal-1995, Close-etal-1997}.
This envelope has a C-shaped structure, probably due to an outflow cavity, as seen from observations with the Hubble Space Telescope \citep{Stapelfeldt-etal-1995}, the Subaru Telescope \citep{Murakawa-etal-2008}, or recently with the James Webb Space Telescope \citep{Mullin-etal-2024}.
Consequently, we include a stellar envelope in our model so that there is additional material along the line of sight that becomes optically thin from \SI{53}{\um} to \SI{216}{\um}.
Thus, in bands A, C, and D, emission of aligned dust grains of the disk plus envelope is observed, whereas in band E, the envelope is transparent and emission and self-scattering in the disk is observed.

\subsection{Model of HL Tau and its circumstellar material}
\label{subsec:disk-env-model}

\begin{table}

    \centering
    \caption{
        General parameters of the model of HL~Tau and its circumstellar material.
        See Sect.~\ref{subsec:disk-env-model} to~\ref{subsec:bfield-model} for details.
    }
    \label{tab:model-parameter}

    \renewcommand{\arraystretch}{1.2}
    \small
    \begin{tabular*}{\linewidth}{@{\extracolsep{\fill}} l r}
            \hline\hline
        Parameter                                         & Value \\
            \hline
            \multicolumn{2}{c}{{Stellar parameters}} \\
        Stellar luminosity                                & \SI{11}{L_\odot} \\
        Effective temperature                             & \SI{4000}{K} \\
        Stellar radius                                    & \SI{6.9}{R_\odot} \\
            \multicolumn{2}{c}{{Disk parameters}} \\
        Inclination                                       & \ang{46.7} \\
        Position angle                                    & \ang{138} \\
        Reference disk density $\rho_0^\mathrm{disk}$     & \SI{7.2e-14}{g.cm^{-3}} \\
        Characteristic disk radius $R_\mathrm{c}$         & \SI{100}{au} \\
        Reference scale height $h_0$                      & \SI{15}{au} \\
        Radial disk profile $\alpha$                      & \num{0.95} \\
        Disk flaring $\beta$                              & \num{1.15} \\
        Inner radius                                      & \SI{0.4}{au} \\
        Resulting disk gas mass                           & \SI{0.13}{M_\odot} \\
            \multicolumn{2}{c}{{Envelope parameters}} \\
        Reference envelope density $\rho_0^\mathrm{env}$  & \SI{2.4e-18}{g.cm^{-3}} \\
        Reference envelope radius $R_\mathrm{e}$          & \SI{1000}{au} \\
        Radial envelope profile $\gamma$                  & \num{0.65} \\
        Outer radius                                      & \SI{e4}{au} \\
        Resulting envelope gas mass $M_\mathrm{env}$      & \SI{0.016}{M_\odot} \\
        Magnetic field strength                           & \SI{1}{mG} \\
        Inner magnetic field radius                       & \SI{100}{au} \\
        Magnetic field position angle                     & \ang{25} \\
            \multicolumn{2}{c}{{Dust parameters}} \\
        Minimum grain size                                & \SI{0.01}{\um} \\
        Maximum grain size $a_\mathrm{max}$ (disk)        & \SI{35}{\um} \\
        Distribution exponent $q$ (disk)                  & \num{3.5} \\
        Maximum grain size $a_\mathrm{max}$ (envelope)    & \SI{1}{\um} \\
        Distribution exponent $q$ (envelope)              & \num{4.0} \\
        Axis ratio of spheroids                           & $2/3$ \\
        Gas-to-dust mass ratio                            & \num{100} \\
        Reference RAT efficiency $Q_0$                    & \num{0.4} \\
        RAT efficiency exponent $\eta$                    & \num{3} \\
        Rayleigh reduction factor                         & \num{0.3} \\
            \hline
    \end{tabular*}

\end{table}

In Table~\ref{tab:model-parameter}, the general model parameters are summarized.
The central radiation source, HL~Tau, is assumed to have a stellar luminosity of \SI{11}{L_\odot} and an effective temperature of \SI{4000}{K} \citep{Menshchikov-etal-1999, Pinte-etal-2016, Liu-etal-2017}.

For the circumstellar disk, we applied a viscous accretion disk model \citep{Lynden-Bell-Pringle-1974, Pringle-1981, Hartmann-etal-1998, Andrews-etal-2009}.
Here, the radial profile is described by a power law tapered by an exponential function.
Using cylindrical coordinates ($r$, $z$), the density distribution is defined by
\begin{equation}
    \label{eq:disk-dust-density}
    \rho_\mathrm{disk} = \rho_0^\mathrm{disk} \left( \frac{r}{R_\mathrm{c}} \right)^{-\alpha}
        \exp \left[ -\left( \frac{r}{R_\mathrm{c}} \right)^{2 - \alpha + \beta} \right]
        \exp \left[ -\frac{1}{2} \left( \frac{z}{h(r)} \right)^2 \right]
\end{equation}
with a scale height of
\begin{equation}
    h(r) = h_0 \left( \frac{r}{R_\mathrm{c}} \right)^\beta.
\end{equation}

We fixed the radial profile and the flaring of the disk at $\alpha = \num{0.95}$ and $\beta = \num{1.15}$, respectively, corresponding to the surface density gradient%
\footnote{%
    Using the notation of Eq.~\eqref{eq:disk-dust-density}, the surface density gradient is defined as $\alpha - \beta$.
}
of \num{-0.2} found by CARMA observations \citep{Kwon-etal-2011, Kwon-etal-2015}.
In addition, we set the characteristic disk radius to $R_\mathrm{c} = \SI{100}{au}$ and the reference scale height to $h_0 = \SI{15}{au}$.
These values are consistent with values found for HL~Tau but also several other circumstellar disks \citep[e.g.,][]{Wolf-etal-2003, Madlener-etal-2012, Kirchschlager-etal-2016, Pinte-etal-2016, Liu-etal-2017}.
However, we note that these parameters cannot be precisely constrained by our observations.
The inner radius of \SI{0.4}{au} of the density distribution is about the sublimation radius of silicates \citep[$\sim$\,\SI{1200}{K};][]{Pollack-etal-1994}.
For the reference disk gas density, we used a value of $\rho_0^\mathrm{disk} = \SI{7.2e-14}{g.cm^{-3}}$.
Thus, we have a total disk gas mass of about \SI{0.13}{M_\odot}, which is in the range of previously determined disk masses from submillimeter and millimeter observations \citep{Guilloteau-etal-2011, Kwon-etal-2015, Carrasco-Gonzalez-etal-2016} and fitting of the spectral energy distribution \citep{Robitaille-etal-2007}.

\begin{figure}
    \centering
    \includegraphics[width=\linewidth]{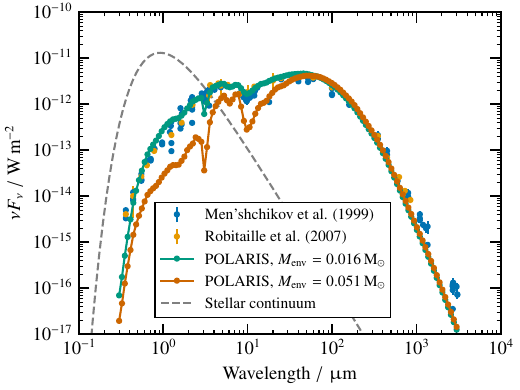}
    \caption{%
        Spectral energy distribution of HL~Tau and our model with two different envelope masses (see Sect.~\ref{subsec:disk-env-model}).
        Photometric data points are from \citet{Menshchikov-etal-1999} and \citet{Robitaille-etal-2007}.
        The stellar continuum is a blackbody with stellar parameters from Table~\ref{tab:model-parameter}.}
    \label{fig:sed-hl-tau}
\end{figure}

In addition to the circumstellar disk, we included a surrounding envelope.
The density distribution of the envelope is described only by a radial gradient,
\begin{equation}
    \rho_\mathrm{env} = \rho_0^\mathrm{env} \left( \frac{R}{R_\mathrm{e}} \right)^{-\gamma}
        \exp \left[ -\left( \frac{R}{R_\mathrm{e}} \right)^{2} \right],
\end{equation}
where $R$ is the radius in spherical coordinates.
We assumed conical polar cavities with an opening angle of \ang{90} which are dust-free, therefore, if $\arccos(\vert z \vert / R) < \pi / 4$, $\rho_\mathrm{env} = 0$.
By setting the reference envelope gas density to $\rho_0^\mathrm{env} = \SI{2.4e-18}{g.cm^{-3}}$ and applying a value of $\gamma = \num{0.65}$, the resulting spectral energy distribution fits best in the near- to mid-infrared wavelength region (see Fig.~\ref{fig:sed-hl-tau}).
With this model, we obtain a total envelope gas mass of about \SI{0.016}{M_\odot}.
For an increasing reference envelope gas density and thus an increasing total envelope gas mass, the resulting spectral energy distribution underestimates the flux at optical to mid-infrared wavelengths (see Fig.~\ref{fig:sed-hl-tau}).
Finally, a reference envelope radius $R_\mathrm{e} = \SI{1000}{au}$ ensures an exponential taper in the outer regions, thus providing an approximate extent of the system determined by millimeter observations \citep{Hayashi-etal-1993}.

To ensure a smooth transition, we combined the two density distributions in such a way that
\begin{equation}
    \rho =
    \begin{cases}
        \rho_\mathrm{disk} & \text{if}\ \rho_\mathrm{disk} > \rho_\mathrm{env}, \\
        \rho_\mathrm{env}  & \text{otherwise}.
    \end{cases}
\end{equation}
The outer radius of the model space was set to \SI{e4}{au}, large enough so that the flux becomes negligible in the outer regions of our model space.
Finally, we assumed a constant gas-to-dust mass ratio of \num{100} throughout the model space.

\subsection{Dust grain properties}
\label{subsec:dust-model}

The dust grains are assumed to be compact homogeneous spheres, consisting of the DSHARP dust composition \citep{Birnstiel-etal-2018} with mass fractions of \SI{47}{\percent} refractory organics \citep{Henning-Stognienko-1996}, \SI{39}{\percent} astronomical silicate \citep{Draine-2003}, \SI{9}{\percent} troilite \citep{Henning-Stognienko-1996}, and \SI{5}{\percent} water ice \citep{Warren-Brandt-2008}.
The composition was mixed using the formula of \citet{Bruggeman-1935}, and the resulting material density is \SI{1.98}{g.cm^{-3}}.
Applying Mie scattering theory \citep{Mie-1908, Bohren-Huffman-1983}, the wavelength-dependent scattering and absorption cross-sections as well as the scattering matrix are calculated using the code MIEX by \citet{Wolf-Voshchinnikov-2004, Wolf-Voshchinnikov-2018}, which is already implemented in POLARIS.

The abundance of dust grain sizes is characterized by a power law,
\begin{equation}
    \label{eq:grain-size}
    n(a) \propto a^{-q}.
\end{equation}
The minimum grain size of the distribution was set to \SI{0.01}{\um}, similar to what is found in the interstellar matter \citep{Mathis-etal-1977}.
For dust particles in the disk, we adopted a value of $q = \num{3.5}$ for the exponent of the distribution.
The maximum grain size $a_\mathrm{max}$ is very sensitive to the scattering efficiency and thus to polarization \citep[e.g.,][]{Kataoka-etal-2015, Brunngraber-Wolf-2019}.

\begin{figure}
    \centering
    \includegraphics[width=\linewidth]{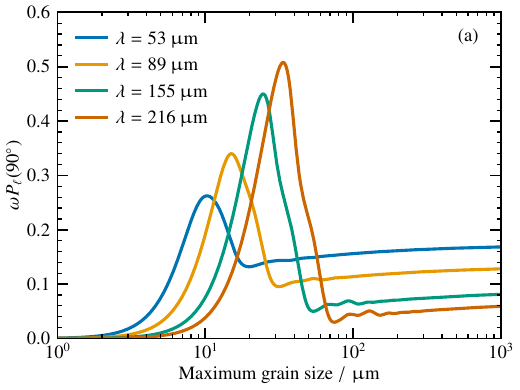}
    \includegraphics[width=\linewidth]{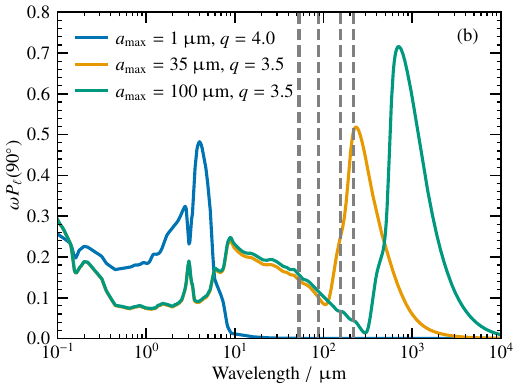}
    \caption{
        Single scattering albedo $\omega$ times the single scattering polarization degree $P_\ell$ at a scattering angle of \ang{90} as a function of maximum grain size ({top}) and wavelength ({bottom}).
        In the top figure (a), the different line colors represent the four central wavelengths of the SOFIA/HAWC+ bands A, C, D, and E.
        The size distribution is described by Eq.~\eqref{eq:grain-size} with an exponent of $q = \num{3.5}$.
        In the bottom figure (b), the different line colors represent the case of different maximum grain sizes $a_\mathrm{max}$ and exponents $q$ of the size distribution.
        The gray dashed vertical lines indicate the central wavelength of the SOFIA/HAWC+ bands A, C, D, and E.
        See Sect.~\ref{subsec:dust-model} for details.
    }
    \label{fig:dust-polarization}
\end{figure}

Since we propose scattering-induced polarization at a wavelength of \SI{216}{\um}, the contribution of polarization due to scattering must exceed the polarization due to emission, and thus the scattering efficiency should be large enough.
Following the approach of \citet{Kataoka-etal-2015}, Fig.~\ref{fig:dust-polarization}a shows the single scattering albedo times the degree of polarization of single scattered radiation at a scattering angle of \ang{90} as a function of maximum grain size.
For the assumed grain properties, the highest degree of polarization is found for a maximum grain size of \SI{35}{\um}.
Since the protoplanetary disk is inclined with respect to the observer, there are not only exact scattering angles of \ang{90}.
However, the highest degree of polarization was consistently found around a maximum grain size of \SI{35}{\um} for a wavelength of \SI{216}{\um}, for example, for scattering angles of \ang{45} or \ang{135}.

\begin{figure}
    \centering
    \includegraphics[width=\linewidth]{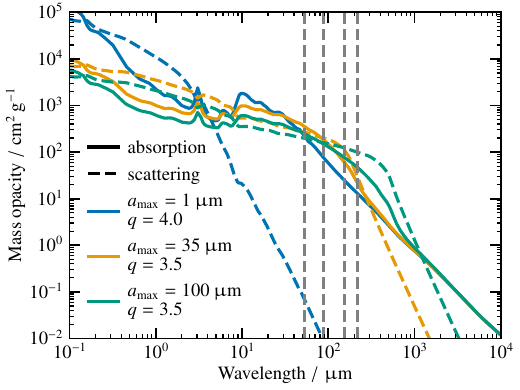}
    \caption{
        Absorption mass opacity ({solid line}) and scattering mass opacity ({dashed line}) as a function of wavelength.
        The different line colors represent different maximum grain sizes $a_\mathrm{max}$ and exponents $q$ of the size distribution (Eq.~\ref{eq:grain-size}).
        The gray dashed vertical lines indicate the central wavelength of the SOFIA/HAWC+ bands A, C, D, and E.
        See Sect.~\ref{subsec:dust-model} for details.
    }
    \label{fig:dust-opacity}
\end{figure}

In contrast, \citet{Kataoka-etal-2017} or \citet{Stephens-etal-2023} found larger grain sizes of up to \SI{100}{\um}.
Despite the scattering mass opacity increases if a maximum grain size of \SI{100}{\um} is assumed (see Fig.~\ref{fig:dust-opacity}), the single scattering albedo times the degree of polarization decreases at a wavelength of \SI{216}{\um} (see Fig.~\ref{fig:dust-polarization}a and~Fig.~\ref{fig:dust-polarization}b).
This is due to the fact that the highest values for $\omega P_\mathrm{\ell}$ are found if $2\pi a_\mathrm{max} / \lambda \approx 1$.
Consequently, the flip in the polarization in band E can not be reproduced by self-scattering if we assume a maximum grain size of \SI{100}{\um}.

This is also consistent with observations of \citet{Li-etal-2018}, since the scattered radiation of these larger grain sizes does not significantly contribute to the mid-infrared polarization (see Fig.~\ref{fig:dust-polarization}b).
In particular, in this wavelength range, the observed radiation is mainly caused by smaller grains located in higher disk regions \citep{Andrews-2020}.
Moreover, with increasing optical depth at mid- to near-infrared wavelengths, the fraction of multiple scattered radiation increases, decreasing the degree of polarization.
Thus, the emission or absorption of dust grains dominates the net polarization, while dichroic absorption of aligned grains in the optically thick envelope would be the major source of near- and mid-infrared polarization \citep{Li-etal-2018}.
We note that \citet{Li-etal-2018} use a slightly different dust composition compared to this study.

Finally, in the envelope, much smaller grains were assumed with a maximum size of \SI{1}{\um} and an exponent of $q = \num{4.0}$.
The steeper grain size distribution can better reproduce the profile of the SED in the near-infrared and optical wavelength region.

Adopting this dust mixture, we obtain a total optical depth of envelope and disk toward the central star of about \num{7.6} at a wavelength of \SI{1.25}{\um} that is in agreement with observations by \citet{Close-etal-1997} and results by \citet{Lucas-etal-2004}.
This supports the hypothesis that in the mid-infrared, we mainly observe smaller particles in the upper layers of the disk and in the envelope \citep{Menshchikov-etal-1999, Robitaille-etal-2007, Kwon-etal-2011}.
In addition, the opacity of the dust in the envelope decreases strongly between \SI{155}{\um} and \SI{216}{\um} (see Fig.~\ref{fig:dust-opacity}).
Consequently, the total optical depth from the central radiation source along the line of sight decreases from \num{0.16} at a wavelength of \SI{53}{\um} to about \num{0.01} at a wavelength of \SI{216}{\um}.
Thus, in band E, the envelope becomes transparent and the emitted as well as scattered radiation inside the disk dominates the total flux.
In particular, polarization in the submillimeter wavelength region arises from self-scattering at larger grains in the circumstellar disk \citep{Kataoka-etal-2016a, Kataoka-etal-2017}.
This finding is also in agreement with the theory of dust settling \citep{Dubrulle-etal-1995, Brunngraber-Wolf-2020, Ueda-etal-2021}.

Non-spherical particles can align, causing polarization due to dichroic emission or absorption.
For this purpose, we assumed oblate dust grains.
Applying the discrete-dipole approximation \citep{Draine-Flatau-1994}, the cross-sections were calculated using the code DDSCAT 7.3.3 \citep{Draine-Flatau-2000, Draine-Flatau-2008}.\footnote{%
    We used $N = \num{281250}$ dipoles, which corresponds to an upper limit of $a / \lambda \lesssim \num{2.6}$ for calculations of cross-sections \citep{Draine-Flatau-2013}.
}
The ratio of minor to major axis of the oblate spheroids was set to $2/3$ \citep{Hildebrand-Dragovan-1995}.
Similarly to spherical dust grains, the optical properties of spheroidal dust grains were calculated assuming the size distribution defined by Eq.~\eqref{eq:grain-size} for the corresponding equal-volume sphere radii.

In the case of the RAT mechanism, a fundamental quantity, which describes the torque on grains and whether a grain is aligned, is the RAT efficiency $Q$ \citep{Draine-Weingartner-1996, Draine-Weingartner-1997}.
It is a function of grain size $a$ and wavelength $\lambda$, and is approximated by
\begin{equation}
    Q =
    \begin{cases}
        Q_0 & \text{if } \lambda \leq \num{1.8} a, \\
        Q_0 \left( \frac{\lambda}{\num{1.8} a} \right)^{-\eta} & \text{otherwise},
    \end{cases}
\end{equation}
where the parameters $Q_0$ and $\eta$ depend on the shape of the grain and grain material, however, are poorly constrained.
Therefore, we apply the average values of $Q_0 = \num{0.4}$ and $\eta = \num{3}$ in this study \citep[see, e.g.,][]{Lazarian-Hoang-2007a, Hoang-Lazarian-2014, Herranen-etal-2019, Reissl-etal-2020}.
Based on the ambient magnetic field, radiation field, gas density and gas temperature, POLARIS calculates the minimum and maximum alignment radius.

Furthermore, several effects, such as imperfectly aligned grains, reduce the net polarization, which can be approximated by the Rayleigh reduction factor \citep{Greenberg-1968, Lee-Draine-1985, Roberge-Lazarian-1999}.
However, the decrease in the degree of polarization by considering imperfect alignment can be compensated by increasing the axis ratio of the grains, and these values cannot be determined by our observations.
Thus, we assumed that grains -- if they are aligned -- are imperfectly aligned to the magnetic field with a Rayleigh reduction factor of \num{0.3} while keeping the axis ratio of the dust grains constant at $2/3$ throughout the model space.

\subsection{Magnetic field properties}
\label{subsec:bfield-model}

Finally, the model is permeated by a magnetic field.
In this study, we assume that the polarized emission arises solely from the surrounding envelope, while polarization in the disk is due to self-scattering only.
As already mentioned in Sect.~\ref{subsec:mir-mm-polarization}, it is unlikely that RAT alignment occurs in the disk of HL~Tau.
Only dust grains located at $R > \SI{100}{au}$ were able to align with the magnetic field in our model.

In this study, we apply a uniform magnetic field with a strength of $B = \SI{1}{mG}$ throughout the model space.
This value is lower than what is measured in younger protostellar cores, high-mass star formation regions \citep{Crutcher-2012, Hull-Zhang-2019}, or the upper limit in the circumstellar disk AS~209 \citep{Harrison-etal-2021}, but larger than what is typically found for the interstellar magnetic field or molecular clouds \citep{Crutcher-1999, Crutcher-2012}.
We note that in our envelope model, the grains are predominantly aligned as a result of RATs, since the magnetic field strength is larger than the estimated critical magnetic field strength \citep{Hughes-etal-2009} based on the dust density and temperature of the envelope.
However, for magnetic field strengths of $\lesssim$\,\SI{0.1}{mG}, the polarization vectors start to rotate at a wavelength of \SI{155}{\um}, which does not fit the observations.
In addition, the power-law relation between the magnetic field strength and density for molecular clouds from theory and observations \citep{Mestel-1966, Crutcher-1999, Crutcher-etal-2010, Crutcher-2012, Dudorov-Khaibrakhmanov-2014} does not have any significant impact on the grain alignment in our model space.
Therefore, only the projected orientation of the magnetic field matters for the radiative transfer simulations.

By assuming that the observed orientation of polarization arises from dichroic emission, the magnetic field direction has to be perpendicular to the polarization vectors.
Thus, we assumed that the direction of the field has a projected position angle of \ang{25}, resulting in a deviation of about \ang{23} to the projected minor axis of the inclined disk.
In addition, all field lines are parallel.
Since only the projected orientation is known, we assume that the magnetic field lines are perpendicular to the line of sight.
Thus, the polarization caused by emission is the largest.
In general, by decreasing the angle between the observer and the magnetic field lines, the degree of polarization decreases because the short axis of the oblate spheroids becomes parallel to the magnetic field lines.

\subsection{Simulation procedure}
\label{subsec:sim-procedure}

\begin{figure*}

    \begin{center}

    \begin{minipage}{0.49\linewidth}
        \centering
        \includegraphics[width=\linewidth]{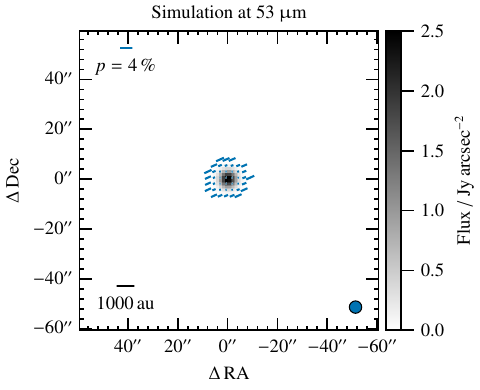}
    \end{minipage}
    \hfill
    \begin{minipage}{0.49\linewidth}
        \centering
        \includegraphics[width=\linewidth]{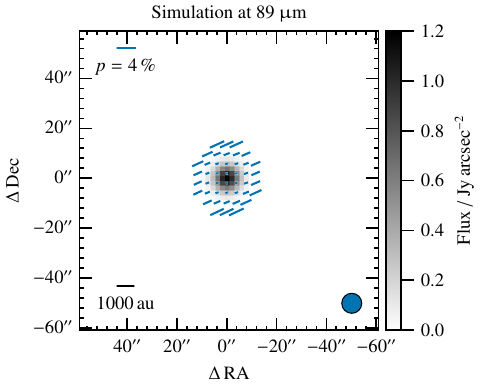}
    \end{minipage}

    \vspace*{0.8em}

    \begin{minipage}{0.49\linewidth}
        \centering
        \includegraphics[width=\linewidth]{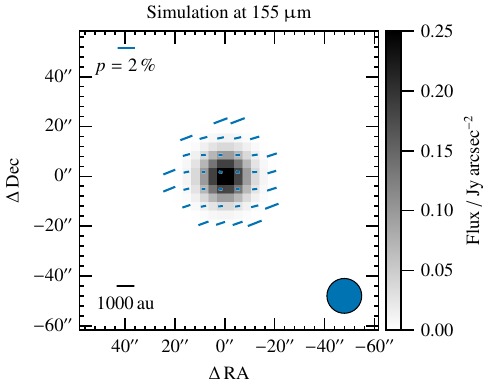}
    \end{minipage}
    \hfill
    \begin{minipage}{0.49\linewidth}
        \centering
        \includegraphics[width=\linewidth]{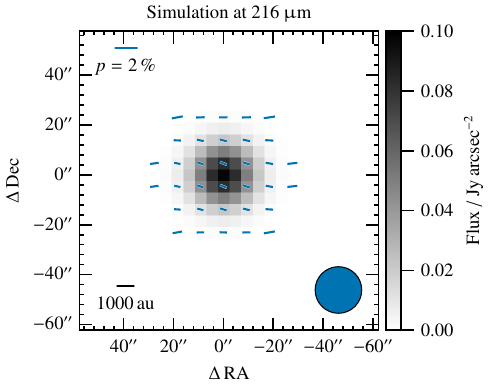}
    \end{minipage}

    \end{center}

    \caption{
        Simulated polarization maps based on the model described in Sect.~\ref{subsec:disk-env-model} at \SI{53}{\um} ({top left}), \SI{89}{\um} ({top right}), \SI{155}{\um} ({bottom left}), and \SI{216}{\um} ({bottom right}).
        The surface brightness (gray scale) is given in units of \si{Jy.arcsec^{-2}} and is overlaid with polarization vectors in blue.
        The length and orientation of the vectors give the degree and angle of polarization, respectively.
        In contrast to the observational criteria, only vectors where $I / I_\mathrm{max} > \num{e-3}$ and $p > \SI{0.1}{\percent}$ are shown.
        The beam size (FWHM) at each corresponding SOFIA/HAWC+ wavelength band is indicated in the respective lower right corner.
        The scale bar (\SI{1000}{au}) corresponds to an assumed distance of HL~Tau of \SI{140}{pc}.
        The images are cropped to an image size of \SI{2}{arcmin} $\times$ \SI{2}{arcmin}.
        See Sect.~\ref{subsec:sim-results} for details.
    }
    \label{fig:polaris-polmap}

\end{figure*}

First, POLARIS calculated the dust temperature and the minimum as well as the maximum alignment radius of the dust grains.
Here, external torques due to radiation or damping due to collisions with gas atoms determine whether a grain is aligned with the magnetic field \citep{Hoang-Lazarian-2014}.
Next, it computed the resulting Stokes parameters assuming spherical particles.
Therein, the unpolarized direct stellar and thermally reemitted radiation as well as the polarized scattered radiation of the star and the dust, that is self-scattering, are considered.
Subsequently, the spherical grains were replaced with oblate shaped particles that are aligned to the magnetic field if their radii are above and below the minimum and maximum alignment radius, respectively.
Finally, the polarized radiation due to dichroic emission and absorption was calculated and added to the unpolarized radiation emitted from the previous simulation step.

\subsection{Simulation results}
\label{subsec:sim-results}

\begin{figure*}

    \begin{center}

    \begin{minipage}{0.49\linewidth}
        \centering
        \includegraphics[width=\linewidth]{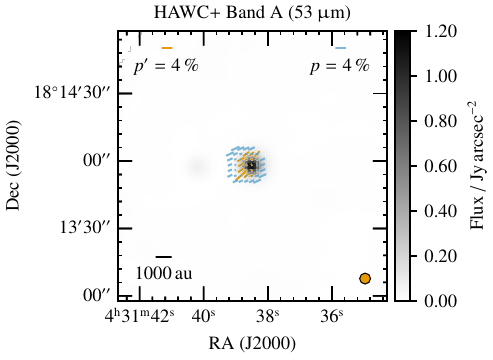}
    \end{minipage}
    \hfill
    \begin{minipage}{0.49\linewidth}
        \centering
        \includegraphics[width=\linewidth]{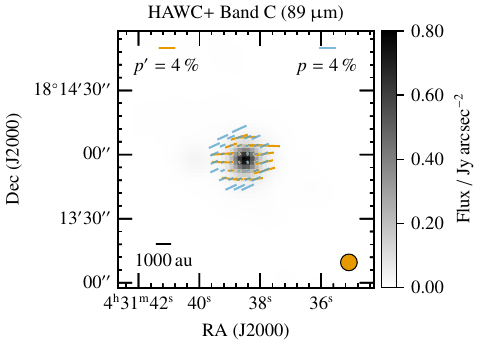}
    \end{minipage}

    \vspace*{0.8em}

    \begin{minipage}{0.49\linewidth}
        \centering
        \includegraphics[width=\linewidth]{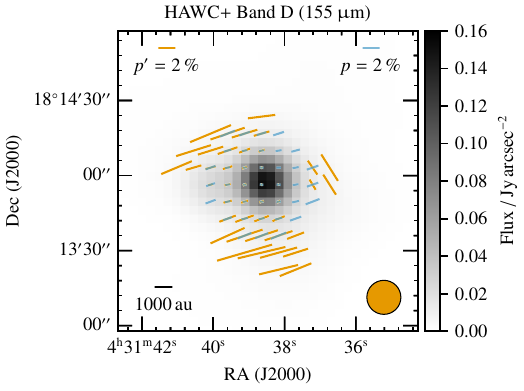}
    \end{minipage}
    \hfill
    \begin{minipage}{0.49\linewidth}
        \centering
        \includegraphics[width=\linewidth]{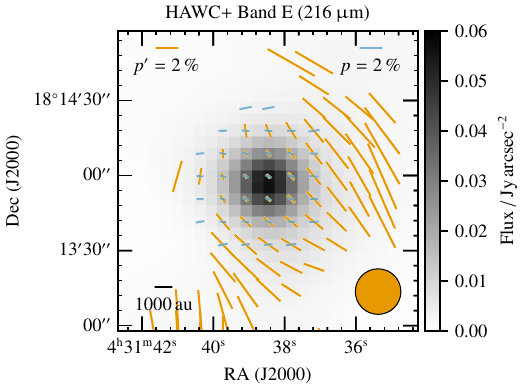}
    \end{minipage}

    \end{center}

    \caption{
        Similar to Fig.~\ref{fig:sofia-polmap}, but overlaid with polarization vectors from the simulations in blue (Fig.~\ref{fig:polaris-polmap}).
        See Sect.~\ref{subsec:sim-results} for details.
    }
    \label{fig:sofia-polaris-polmap}

\end{figure*}

Figure~\ref{fig:polaris-polmap} shows the resulting polarization maps from the numerical simulations.
The maps have the same pixel scale and are convolved by a Gaussian beam with the respective SOFIA/HAWC+ beam size of the corresponding wavelength.
Similarly to the observations presented in Sect.~\ref{sec:results}, the surface brightness is overlaid with polarization vectors and the images are cropped to an image size of \SI{2}{arcmin} $\times$ \SI{2}{arcmin}.
In general, the total flux decreases with increasing wavelength.
However, the flux of the simulation exceeds the flux of the observations, especially in band A, by a factor of about \num{2}.
Additionally, the degree of polarization of the simulations is somewhat lower compared to the observations at the respective SOFIA/HAWC+ wavelength bands.

At a wavelength of \SI{53}{\um}, \SI{89}{\um}, and \SI{155}{\um}, the orientation of the polarization vectors has a position angle of about \ang{110}.
Furthermore, at a wavelength of \SI{216}{\um}, the polarization orientation changes to a direction that is somewhat parallel to the minor axis of the inclined disk projection.
As a result, we are able to reproduce the flip in the polarization.
In particular, aligned non-spherical dust grains cause polarization at these shorter wavelengths, while self-scattering causes polarization at \SI{216}{\um}.

Figure~\ref{fig:sofia-polaris-polmap} shows the SOFIA/HAWC+ polarization maps overlaid with polarization vectors from the simulations at the respective central wavelength.
The exact position angle of the polarization could not be reproduced in all wavelength bands.
For example, we did not reproduce the elliptically shaped polarization pattern in band C.
One possible reason for this deviation is that the uniform magnetic field considered in our model is too simple and the underlying magnetic field morphology has a more complex structure.
For wavelengths of \SI{53}{\um}, \SI{89}{\um}, and \SI{155}{\um}, the degree of polarization decreases toward the inner brighter region, which is consistent with the observations.
Moreover, at a wavelength of \SI{216}{\um}, the polarization in the central region is in agreement with the observation with a value of approximately \SI{0.5}{\percent}.

In contrast, the polarization orientation of band E is not well fitted, although we could reproduce the orientation of the polarization arising from self-scattering.
One possible reason is that polarization due to scattered radiation is not completely dominant at this wavelength, whereas polarization due to emitted radiation of aligned dust grains is still present.
Thus, either a decrease in polarization due to emission of aligned grains or an increase in polarization due to self-scattering would solve the problem.
In the first case, the polarization arising from dichroic emission decreases with increasing grain porosity \citep{Kirchschlager-etal-2019}.
However, this would also have an impact on the polarization orientation at a wavelength of \SI{155}{\um} (band D).
On the other hand, scattering of elongated aligned dust grains produces higher polarization compared to compact spheres \citep{Kirchschlager-Bertrang-2020}.
In addition, porous dust aggregates show a higher degree of polarization \citep{Tazaki-etal-2019b} or broaden the possible range for the maximum grain size \citep{Tazaki-etal-2019a} compared to compact dust grains.
Finally, the degree of polarization of large irregular grains derived from laboratory measurements differs from the polarization derived from the Mie theory, where compact spheres are considered \citep{Lin-etal-2023}.
Nevertheless, these non-ideal effects are beyond the scope of this study.

Finally, the extended polarization pattern and the degree of polarization in the outer regions observed in band E are not reproduced as well.
Therefore, it is uncertain whether the observed polarization can be attributed to HL~Tau, since the envelope is optically thin at this wavelength.
However, if the polarization is caused by surrounding dust, which is not considered in our model, then the elliptical orientation of the polarization vectors is probably a combination of two effects.
First, the circumstellar disk appears as a point-like source, and secondly, the beam convolution does not play any role at that distance to the central source.

%%%%%%%%%%%%%%%%%%%%%%%%%%%%%%%%%%%%%%%%%%%%%%%%%%%%%%%%%%%%%%

\section{Conclusions}
\label{sec:conclusions}

We present the first polarimetric observations of a circumstellar disk in the far-infrared wavelength region.
In particular, we report flux and linear polarization measurements for HL~Tau, a young stellar object surrounded by a circumstellar disk, in the SOFIA/HAWC+ bands A (\SI{53}{\um}), C (\SI{89}{\um}), D (\SI{155}{\um}), and E (\SI{216}{\um}).
Although the size of the disk around HL~Tau is smaller than the pixel scale of the observations, the origin of polarization of the polarization maps can be attributed to the dust in the protoplanetary disk around HL~Tau.

The orientation of the polarization vectors is roughly parallel to that of the major axis of the projection of the inclined disk in the bands A, C, and D.
We conclude that the polarization at these wavelengths is caused by aligned non-spherical dust grains.
As a result, we are able to constrain the projected direction of the magnetic field.
The position angle of the magnetic field lines is about \ang{25}, and thus close to the minor axis of the disk at a position angle of \ang{48}.
In contrast, the polarization vectors are parallel to the disk minor axis in band E.
Here, the polarization pattern is consistent with that of self-scattering.

Performing three-dimensional Monte Carlo polarized radiative transfer simulations with POLARIS, we were able to reproduce the flip of the polarization vectors with increasing wavelength.
We modeled a viscous accretion disk combined with an envelope whose density distribution is described by a power law.
By assuming the commonly used DSHARP dust composition, we could constrain the maximum grain size of \SI{35}{\um} to reproduce the scattering-induced polarization at a wavelength of \SI{216}{\um}.
However, we were unable to fully match the polarization pattern in band E.
At shorter wavelengths, polarized emission is caused by oblate spheroids that are aligned to a magnetic field with a position angle of about \ang{25}.
In particular, polarization in bands A, C, and D cannot be attributed to dichroic extinction.
Moreover, we conclude that the polarized emission and the absence of scattering can not be used as an explanation for the underlying mechanism that causes the flip in the orientation of the polarization in band E.
This is because the optical depth of our envelope model is too small in the far-infrared wavelength region ($\sim$\,\num{0.16} at \SI{53}{\um}).

There are still deviations in the total flux, the degree of polarization, and the orientation of the polarization vectors.
Most importantly, the simple envelope with conical polar cavities does not include dust surrounding the system on larger scales, nor does it take into account the molecular outflow of HL~Tau.
Furthermore, our model does not consider the neighboring system XZ~Tau, which might also contribute to the total flux and polarization due to the relatively large beam size of SOFIA/HAWC+ compared to the apparent separation of both systems.

In conclusion, these far-infrared polarimetric observations not only give unique insights into the magnetic field around HL~Tau, but also provide complementary constraints for the spatial distribution of the dust and dust properties if compared to flux measurements alone.

%%%%%%%%%%%%%%%%%%%%%%%%%%%%%%%%%%%%%%%%%%%%%%%%%%%%%%%%%%%%%%

\section*{Data availability}

The data obtained with SOFIA/HAWC+ and used in this study are available in the SOFIA Science Data Archive at the Infrared Science Archive (IRSA) under Plan ID
\href{https://irsa.ipac.caltech.edu/applications/sofia/?api=search&spatialConstraints=allsky&planId=09_0084&&instrument=HAWC_PLUS&processingLevel=LEVEL_4&execute=true}{09\_0084}.

%%%%%%%%%%%%%%%%%%%%%%%%%%%%%%%%%%%%%%%%%%%%%%%%%%%%%%%%%%%%%%

\begin{acknowledgements}
    The authors thank the anonymous referee for useful and constructive comments and suggestions.
    Based on observations made with the NASA/DLR Stratospheric Observatory for Infrared Astronomy (SOFIA).
    SOFIA is jointly operated by the Universities Space Research Association, Inc. (USRA), under NASA contract NNA17BF53C, and the Deutsches SOFIA Institut (DSI) under DLR contract 50 OK 2002 to the University of Stuttgart.
    This research was funded through DLR/BMWK grant 50 OR 2210.
    The publication made use of
    \href{https://github.com/aplpy/aplpy}{APLpy} \citep{Robitaille-Bressert-2012},
    \href{https://github.com/astropy/astropy}{Astropy} \citep{Astropy-etal-2013, Astropy-etal-2018, Astropy-etal-2022},
    \href{https://github.com/matplotlib/matplotlib}{Matplotlib} \citep{Hunter-2007},
    \href{https://github.com/numpy/numpy}{NumPy} \citep{Harris-etal-2020},
    the \href{https://ui.adsabs.harvard.edu}{Astrophysics Data System}, funded by NASA under Cooperative Agreement 80NSSC21M00561,
    and the \href{https://irsa.ipac.caltech.edu/}{NASA/IPAC Infrared Science Archive}, funded by the National Aeronautics and Space Administration and operated by the California Institute of Technology.
\end{acknowledgements}

\bibliographystyle{aalink}
\bibliography{bibliography.bib}

\begin{thebibliography}{127}
\expandafter\ifx\csname natexlab\endcsname\relax\def\natexlab#1{#1}\fi

\bibitem[{{ALMA Partnership} {et~al.}(2015){ALMA Partnership}, {Brogan}, {P{\'e}rez}, {Hunter}, {Dent}, {Hales}, {Hills}, {Corder}, {Fomalont}, {Vlahakis}, {Asaki}, {Barkats}, {Hirota}, {Hodge}, {Impellizzeri}, {Kneissl}, {Liuzzo}, {Lucas}, {Marcelino}, {Matsushita}, {Nakanishi}, {Phillips}, {Richards}, {Toledo}, {Aladro}, {Broguiere}, {Cortes}, {Cortes}, {Espada}, {Galarza}, {Garcia-Appadoo}, {Guzman-Ramirez}, {Humphreys}, {Jung}, {Kameno}, {Laing}, {Leon}, {Marconi}, {Mignano}, {Nikolic}, {Nyman}, {Radiszcz}, {Remijan}, {Rod{\'o}n}, {Sawada}, {Takahashi}, {Tilanus}, {Vila Vilaro}, {Watson}, {Wiklind}, {Akiyama}, {Chapillon}, {de Gregorio-Monsalvo}, {Di Francesco}, {Gueth}, {Kawamura}, {Lee}, {Nguyen Luong}, {Mangum}, {Pietu}, {Sanhueza}, {Saigo}, {Takakuwa}, {Ubach}, {van Kempen}, {Wootten}, {Castro-Carrizo}, {Francke}, {Gallardo}, {Garcia}, {Gonzalez}, {Hill}, {Kaminski}, {Kurono}, {Liu}, {Lopez}, {Morales}, {Plarre}, {Schieven}, {Testi}, {Videla}, {Villard}, {Andreani}, {Hibbard}, \&
  {Tatematsu}}]{ALMA-etal-2015}
{ALMA Partnership}, {Brogan}, C.~L., {P{\'e}rez}, L.~M., {et~al.} 2015, \href{https://doi.org/10.1088/2041-8205/808/1/L3}{\textcolor{violet}{\apjl}}, \href{https://ui.adsabs.harvard.edu/abs/2015ApJ...808L...3A}{808, L3}

\bibitem[{{Andersson} {et~al.}(2015){Andersson}, {Lazarian}, \& {Vaillancourt}}]{Andersson-etal-2015}
{Andersson}, B.~G., {Lazarian}, A., \& {Vaillancourt}, J.~E. 2015, \href{https://doi.org/10.1146/annurev-astro-082214-122414}{\textcolor{violet}{\araa}}, \href{https://ui.adsabs.harvard.edu/abs/2015ARA&A..53..501A}{53, 501}

\bibitem[{{Andrews}(2020)}]{Andrews-2020}
{Andrews}, S.~M. 2020, \href{https://doi.org/10.1146/annurev-astro-031220-010302}{\textcolor{violet}{\araa}}, \href{https://ui.adsabs.harvard.edu/abs/2020ARA&A..58..483A}{58, 483}

\bibitem[{{Andrews} {et~al.}(2009){Andrews}, {Wilner}, {Hughes}, {Qi}, \& {Dullemond}}]{Andrews-etal-2009}
{Andrews}, S.~M., {Wilner}, D.~J., {Hughes}, A.~M., {Qi}, C., \& {Dullemond}, C.~P. 2009, \href{https://doi.org/10.1088/0004-637X/700/2/1502}{\textcolor{violet}{\apj}}, \href{https://ui.adsabs.harvard.edu/abs/2009ApJ...700.1502A}{700, 1502}

\bibitem[{{Astropy Collaboration} {et~al.}(2022){Astropy Collaboration}, {Price-Whelan}, {Lim}, {Earl}, {Starkman}, {Bradley}, {Shupe}, {Patil}, {Corrales}, {Brasseur}, {N{\"o}the}, {Donath}, {Tollerud}, {Morris}, {Ginsburg}, {Vaher}, {Weaver}, {Tocknell}, {Jamieson}, {van Kerkwijk}, {Robitaille}, {Merry}, {Bachetti}, {G{\"u}nther}, {Aldcroft}, {Alvarado-Montes}, {Archibald}, {B{\'o}di}, {Bapat}, {Barentsen}, {Baz{\'a}n}, {Biswas}, {Boquien}, {Burke}, {Cara}, {Cara}, {Conroy}, {Conseil}, {Craig}, {Cross}, {Cruz}, {D'Eugenio}, {Dencheva}, {Devillepoix}, {Dietrich}, {Eigenbrot}, {Erben}, {Ferreira}, {Foreman-Mackey}, {Fox}, {Freij}, {Garg}, {Geda}, {Glattly}, {Gondhalekar}, {Gordon}, {Grant}, {Greenfield}, {Groener}, {Guest}, {Gurovich}, {Handberg}, {Hart}, {Hatfield-Dodds}, {Homeier}, {Hosseinzadeh}, {Jenness}, {Jones}, {Joseph}, {Kalmbach}, {Karamehmetoglu}, {Ka{\l}uszy{\'n}ski}, {Kelley}, {Kern}, {Kerzendorf}, {Koch}, {Kulumani}, {Lee}, {Ly}, {Ma}, {MacBride}, {Maljaars}, {Muna}, {Murphy}, {Norman},
  {O'Steen}, {Oman}, {Pacifici}, {Pascual}, {Pascual-Granado}, {Patil}, {Perren}, {Pickering}, {Rastogi}, {Roulston}, {Ryan}, {Rykoff}, {Sabater}, {Sakurikar}, {Salgado}, {Sanghi}, {Saunders}, {Savchenko}, {Schwardt}, {Seifert-Eckert}, {Shih}, {Jain}, {Shukla}, {Sick}, {Simpson}, {Singanamalla}, {Singer}, {Singhal}, {Sinha}, {Sip{\H{o}}cz}, {Spitler}, {Stansby}, {Streicher}, {{\v{S}}umak}, {Swinbank}, {Taranu}, {Tewary}, {Tremblay}, {de Val-Borro}, {Van Kooten}, {Vasovi{\'c}}, {Verma}, {de Miranda Cardoso}, {Williams}, {Wilson}, {Winkel}, {Wood-Vasey}, {Xue}, {Yoachim}, {Zhang}, {Zonca}, \& {Astropy Project Contributors}}]{Astropy-etal-2022}
{Astropy Collaboration}, {Price-Whelan}, A.~M., {Lim}, P.~L., {et~al.} 2022, \href{https://doi.org/10.3847/1538-4357/ac7c74}{\textcolor{violet}{\apj}}, \href{https://ui.adsabs.harvard.edu/abs/2022ApJ...935..167A}{935, 167}

\bibitem[{{Astropy Collaboration} {et~al.}(2018){Astropy Collaboration}, {Price-Whelan}, {Sip{\H{o}}cz}, {G{\"u}nther}, {Lim}, {Crawford}, {Conseil}, {Shupe}, {Craig}, {Dencheva}, {Ginsburg}, {VanderPlas}, {Bradley}, {P{\'e}rez-Su{\'a}rez}, {de Val-Borro}, {Aldcroft}, {Cruz}, {Robitaille}, {Tollerud}, {Ardelean}, {Babej}, {Bach}, {Bachetti}, {Bakanov}, {Bamford}, {Barentsen}, {Barmby}, {Baumbach}, {Berry}, {Biscani}, {Boquien}, {Bostroem}, {Bouma}, {Brammer}, {Bray}, {Breytenbach}, {Buddelmeijer}, {Burke}, {Calderone}, {Cano Rodr{\'\i}guez}, {Cara}, {Cardoso}, {Cheedella}, {Copin}, {Corrales}, {Crichton}, {D'Avella}, {Deil}, {Depagne}, {Dietrich}, {Donath}, {Droettboom}, {Earl}, {Erben}, {Fabbro}, {Ferreira}, {Finethy}, {Fox}, {Garrison}, {Gibbons}, {Goldstein}, {Gommers}, {Greco}, {Greenfield}, {Groener}, {Grollier}, {Hagen}, {Hirst}, {Homeier}, {Horton}, {Hosseinzadeh}, {Hu}, {Hunkeler}, {Ivezi{\'c}}, {Jain}, {Jenness}, {Kanarek}, {Kendrew}, {Kern}, {Kerzendorf}, {Khvalko}, {King}, {Kirkby}, {Kulkarni},
  {Kumar}, {Lee}, {Lenz}, {Littlefair}, {Ma}, {Macleod}, {Mastropietro}, {McCully}, {Montagnac}, {Morris}, {Mueller}, {Mumford}, {Muna}, {Murphy}, {Nelson}, {Nguyen}, {Ninan}, {N{\"o}the}, {Ogaz}, {Oh}, {Parejko}, {Parley}, {Pascual}, {Patil}, {Patil}, {Plunkett}, {Prochaska}, {Rastogi}, {Reddy Janga}, {Sabater}, {Sakurikar}, {Seifert}, {Sherbert}, {Sherwood-Taylor}, {Shih}, {Sick}, {Silbiger}, {Singanamalla}, {Singer}, {Sladen}, {Sooley}, {Sornarajah}, {Streicher}, {Teuben}, {Thomas}, {Tremblay}, {Turner}, {Terr{\'o}n}, {van Kerkwijk}, {de la Vega}, {Watkins}, {Weaver}, {Whitmore}, {Woillez}, {Zabalza}, \& {Astropy Contributors}}]{Astropy-etal-2018}
{Astropy Collaboration}, {Price-Whelan}, A.~M., {Sip{\H{o}}cz}, B.~M., {et~al.} 2018, \href{https://doi.org/10.3847/1538-3881/aabc4f}{\textcolor{violet}{\aj}}, \href{https://ui.adsabs.harvard.edu/abs/2018AJ....156..123A}{156, 123}

\bibitem[{{Astropy Collaboration} {et~al.}(2013){Astropy Collaboration}, {Robitaille}, {Tollerud}, {Greenfield}, {Droettboom}, {Bray}, {Aldcroft}, {Davis}, {Ginsburg}, {Price-Whelan}, {Kerzendorf}, {Conley}, {Crighton}, {Barbary}, {Muna}, {Ferguson}, {Grollier}, {Parikh}, {Nair}, {Unther}, {Deil}, {Woillez}, {Conseil}, {Kramer}, {Turner}, {Singer}, {Fox}, {Weaver}, {Zabalza}, {Edwards}, {Azalee Bostroem}, {Burke}, {Casey}, {Crawford}, {Dencheva}, {Ely}, {Jenness}, {Labrie}, {Lim}, {Pierfederici}, {Pontzen}, {Ptak}, {Refsdal}, {Servillat}, \& {Streicher}}]{Astropy-etal-2013}
{Astropy Collaboration}, {Robitaille}, T.~P., {Tollerud}, E.~J., {et~al.} 2013, \href{https://doi.org/10.1051/0004-6361/201322068}{\textcolor{violet}{\aap}}, \href{https://ui.adsabs.harvard.edu/abs/2013A&A...558A..33A}{558, A33}

\bibitem[{{Bacciotti} {et~al.}(2018){Bacciotti}, {Girart}, {Padovani}, {Podio}, {Paladino}, {Testi}, {Bianchi}, {Galli}, {Codella}, {Coffey}, {Favre}, \& {Fedele}}]{Bacciotti-etal-2018}
{Bacciotti}, F., {Girart}, J.~M., {Padovani}, M., {et~al.} 2018, \href{https://doi.org/10.3847/2041-8213/aadf87}{\textcolor{violet}{\apjl}}, \href{https://ui.adsabs.harvard.edu/abs/2018ApJ...865L..12B}{865, L12}

\bibitem[{{Barnett}(1915)}]{Barnett-1915}
{Barnett}, S.~J. 1915, \href{https://doi.org/10.1103/PhysRev.6.239}{\textcolor{violet}{Phys. Rev.}}, \href{https://ui.adsabs.harvard.edu/abs/1915PhRv....6..239B}{6, 239}

\bibitem[{{Beckwith} \& {Birk}(1995)}]{Beckwith-Birk-1995}
{Beckwith}, S. V.~W. \& {Birk}, C.~C. 1995, \href{https://doi.org/10.1086/309636}{\textcolor{violet}{\apjl}}, \href{https://ui.adsabs.harvard.edu/abs/1995ApJ...449L..59B}{449, L59}

\bibitem[{{Beckwith} {et~al.}(1990){Beckwith}, {Sargent}, {Chini}, \& {Guesten}}]{Beckwith-etal-1990}
{Beckwith}, S. V.~W., {Sargent}, A.~I., {Chini}, R.~S., \& {Guesten}, R. 1990, \href{https://doi.org/10.1086/115385}{\textcolor{violet}{\aj}}, \href{https://ui.adsabs.harvard.edu/abs/1990AJ.....99..924B}{99, 924}

\bibitem[{{Birnstiel} {et~al.}(2018){Birnstiel}, {Dullemond}, {Zhu}, {Andrews}, {Bai}, {Wilner}, {Carpenter}, {Huang}, {Isella}, {Benisty}, {P{\'e}rez}, \& {Zhang}}]{Birnstiel-etal-2018}
{Birnstiel}, T., {Dullemond}, C.~P., {Zhu}, Z., {et~al.} 2018, \href{https://doi.org/10.3847/2041-8213/aaf743}{\textcolor{violet}{\apjl}}, \href{https://ui.adsabs.harvard.edu/abs/2018ApJ...869L..45B}{869, L45}

\bibitem[{{Bohren} \& {Huffman}(1983)}]{Bohren-Huffman-1983}
{Bohren}, C.~F. \& {Huffman}, D.~R. 1983, \href{https://ui.adsabs.harvard.edu/abs/1983asls.book.....B}{{Absorption and scattering of light by small particles}} (Wiley)

\bibitem[{{Bruggeman}(1935)}]{Bruggeman-1935}
{Bruggeman}, D.~A.~G. 1935, \href{https://doi.org/10.1002/andp.19354160705}{\textcolor{violet}{Ann. Phys.}}, \href{https://ui.adsabs.harvard.edu/abs/1935AnP...416..636B}{416, 636}

\bibitem[{{Brunngr{\"a}ber} \& {Wolf}(2019)}]{Brunngraber-Wolf-2019}
{Brunngr{\"a}ber}, R. \& {Wolf}, S. 2019, \href{https://doi.org/10.1051/0004-6361/201935169}{\textcolor{violet}{\aap}}, \href{https://ui.adsabs.harvard.edu/abs/2019A&A...627L..10B}{627, L10}

\bibitem[{{Brunngr{\"a}ber} \& {Wolf}(2020)}]{Brunngraber-Wolf-2020}
{Brunngr{\"a}ber}, R. \& {Wolf}, S. 2020, \href{https://doi.org/10.1051/0004-6361/202037981}{\textcolor{violet}{\aap}}, \href{https://ui.adsabs.harvard.edu/abs/2020A&A...640A.122B}{640, A122}

\bibitem[{{Brunngr{\"a}ber} \& {Wolf}(2021)}]{Brunngraber-Wolf-2021}
{Brunngr{\"a}ber}, R. \& {Wolf}, S. 2021, \href{https://doi.org/10.1051/0004-6361/202040033}{\textcolor{violet}{\aap}}, \href{https://ui.adsabs.harvard.edu/abs/2021A&A...648A..87B}{648, A87}

\bibitem[{{Carrasco-Gonz{\'a}lez} {et~al.}(2016){Carrasco-Gonz{\'a}lez}, {Henning}, {Chandler}, {Linz}, {P{\'e}rez}, {Rodr{\'\i}guez}, {Galv{\'a}n-Madrid}, {Anglada}, {Birnstiel}, {van Boekel}, {Flock}, {Klahr}, {Macias}, {Menten}, {Osorio}, {Testi}, {Torrelles}, \& {Zhu}}]{Carrasco-Gonzalez-etal-2016}
{Carrasco-Gonz{\'a}lez}, C., {Henning}, T., {Chandler}, C.~J., {et~al.} 2016, \href{https://doi.org/10.3847/2041-8205/821/1/L16}{\textcolor{violet}{\apjl}}, \href{https://ui.adsabs.harvard.edu/abs/2016ApJ...821L..16C}{821, L16}

\bibitem[{{Carrasco-Gonz{\'a}lez} {et~al.}(2019){Carrasco-Gonz{\'a}lez}, {Sierra}, {Flock}, {Zhu}, {Henning}, {Chandler}, {Galv{\'a}n-Madrid}, {Mac{\'\i}as}, {Anglada}, {Linz}, {Osorio}, {Rodr{\'\i}guez}, {Testi}, {Torrelles}, {P{\'e}rez}, \& {Liu}}]{Carrasco-Gonzalez-etal-2019}
{Carrasco-Gonz{\'a}lez}, C., {Sierra}, A., {Flock}, M., {et~al.} 2019, \href{https://doi.org/10.3847/1538-4357/ab3d33}{\textcolor{violet}{\apj}}, \href{https://ui.adsabs.harvard.edu/abs/2019ApJ...883...71C}{883, 71}

\bibitem[{{Cho} \& {Lazarian}(2007)}]{Cho-Lazarian-2007}
{Cho}, J. \& {Lazarian}, A. 2007, \href{https://doi.org/10.1086/521805}{\textcolor{violet}{\apj}}, \href{https://ui.adsabs.harvard.edu/abs/2007ApJ...669.1085C}{669, 1085}

\bibitem[{{Close} {et~al.}(1997){Close}, {Roddier}, {J. Northcott}, {Roddier}, \& {Elon Graves}}]{Close-etal-1997}
{Close}, L.~M., {Roddier}, F., {J. Northcott}, M., {Roddier}, C., \& {Elon Graves}, J. 1997, \href{https://doi.org/10.1086/303813}{\textcolor{violet}{\apj}}, \href{https://ui.adsabs.harvard.edu/abs/1997ApJ...478..766C}{478, 766}

\bibitem[{{Crutcher}(1999)}]{Crutcher-1999}
{Crutcher}, R.~M. 1999, \href{https://doi.org/10.1086/307483}{\textcolor{violet}{\apj}}, \href{https://ui.adsabs.harvard.edu/abs/1999ApJ...520..706C}{520, 706}

\bibitem[{{Crutcher}(2012)}]{Crutcher-2012}
{Crutcher}, R.~M. 2012, \href{https://doi.org/10.1146/annurev-astro-081811-125514}{\textcolor{violet}{\araa}}, \href{https://ui.adsabs.harvard.edu/abs/2012ARA&A..50...29C}{50, 29}

\bibitem[{{Crutcher} {et~al.}(2010){Crutcher}, {Wandelt}, {Heiles}, {Falgarone}, \& {Troland}}]{Crutcher-etal-2010}
{Crutcher}, R.~M., {Wandelt}, B., {Heiles}, C., {Falgarone}, E., \& {Troland}, T.~H. 2010, \href{https://doi.org/10.1088/0004-637X/725/1/466}{\textcolor{violet}{\apj}}, \href{https://ui.adsabs.harvard.edu/abs/2010ApJ...725..466C}{725, 466}

\bibitem[{{Dent} {et~al.}(2019){Dent}, {Pinte}, {Cortes}, {M{\'e}nard}, {Hales}, {Fomalont}, \& {de Gregorio-Monsalvo}}]{Dent-etal-2019}
{Dent}, W.~R.~F., {Pinte}, C., {Cortes}, P.~C., {et~al.} 2019, \href{https://doi.org/10.1093/mnrasl/sly181}{\textcolor{violet}{\mnras}}, \href{https://ui.adsabs.harvard.edu/abs/2019MNRAS.482L..29D}{482, L29}

\bibitem[{{Draine}(2003)}]{Draine-2003}
{Draine}, B.~T. 2003, \href{https://doi.org/10.1146/annurev.astro.41.011802.094840}{\textcolor{violet}{\araa}}, \href{https://ui.adsabs.harvard.edu/abs/2003ARA&A..41..241D}{41, 241}

\bibitem[{{Draine} \& {Flatau}(1994)}]{Draine-Flatau-1994}
{Draine}, B.~T. \& {Flatau}, P.~J. 1994, \href{https://doi.org/10.1364/JOSAA.11.001491}{\textcolor{violet}{J. Opt. Soc. Am. A}}, \href{https://ui.adsabs.harvard.edu/abs/1994JOSAA..11.1491D}{11, 1491}

\bibitem[{{Draine} \& {Flatau}(2000)}]{Draine-Flatau-2000}
{Draine}, B.~T. \& {Flatau}, P.~J. 2000, \href{https://ui.adsabs.harvard.edu/abs/2000ascl.soft08001D}{{DDSCAT: The discrete dipole approximation for scattering and absorption of light by irregular particles}}, \href{https://ascl.net/0008.001}{\textcolor{violet}{ascl:0008.001}}

\bibitem[{{Draine} \& {Flatau}(2008)}]{Draine-Flatau-2008}
{Draine}, B.~T. \& {Flatau}, P.~J. 2008, \href{https://doi.org/10.1364/JOSAA.25.002693}{\textcolor{violet}{J. Opt. Soc. Am. A}}, \href{https://ui.adsabs.harvard.edu/abs/2008JOSAA..25.2693D}{25, 2693}

\bibitem[{{Draine} \& {Flatau}(2013)}]{Draine-Flatau-2013}
{Draine}, B.~T. \& {Flatau}, P.~J. 2013, \href{https://ui.adsabs.harvard.edu/abs/2013arXiv1305.6497D}{{User Guide for the Discrete Dipole Approximation Code DDSCAT 7.3}}, \href{https://arxiv.org/abs/1305.6497}{\textcolor{violet}{arXiv:1305.6497}}

\bibitem[{{Draine} \& {Weingartner}(1996)}]{Draine-Weingartner-1996}
{Draine}, B.~T. \& {Weingartner}, J.~C. 1996, \href{https://doi.org/10.1086/177887}{\textcolor{violet}{\apj}}, \href{https://ui.adsabs.harvard.edu/abs/1996ApJ...470..551D}{470, 551}

\bibitem[{{Draine} \& {Weingartner}(1997)}]{Draine-Weingartner-1997}
{Draine}, B.~T. \& {Weingartner}, J.~C. 1997, \href{https://doi.org/10.1086/304008}{\textcolor{violet}{\apj}}, \href{https://ui.adsabs.harvard.edu/abs/1997ApJ...480..633D}{480, 633}

\bibitem[{{Dubrulle} {et~al.}(1995){Dubrulle}, {Morfill}, \& {Sterzik}}]{Dubrulle-etal-1995}
{Dubrulle}, B., {Morfill}, G., \& {Sterzik}, M. 1995, \href{https://doi.org/10.1006/icar.1995.1058}{\textcolor{violet}{\icarus}}, \href{https://ui.adsabs.harvard.edu/abs/1995Icar..114..237D}{114, 237}

\bibitem[{{Dudorov} \& {Khaibrakhmanov}(2014)}]{Dudorov-Khaibrakhmanov-2014}
{Dudorov}, A.~E. \& {Khaibrakhmanov}, S.~A. 2014, \href{https://doi.org/10.1007/s10509-014-1900-4}{\textcolor{violet}{\apss}}, \href{https://ui.adsabs.harvard.edu/abs/2014Ap&SS.352..103D}{352, 103}

\bibitem[{{Galli} {et~al.}(2018){Galli}, {Loinard}, {Ortiz-L{\'e}on}, {Kounkel}, {Dzib}, {Mioduszewski}, {Rodr{\'\i}guez}, {Hartmann}, {Teixeira}, {Torres}, {Rivera}, {Boden}, {Evans}, {Brice{\~n}o}, {Tobin}, \& {Heyer}}]{Galli-etal-2018}
{Galli}, P. A.~B., {Loinard}, L., {Ortiz-L{\'e}on}, G.~N., {et~al.} 2018, \href{https://doi.org/10.3847/1538-4357/aabf91}{\textcolor{violet}{\apj}}, \href{https://ui.adsabs.harvard.edu/abs/2018ApJ...859...33G}{859, 33}

\bibitem[{{Gold}(1952)}]{Gold-1952}
{Gold}, T. 1952, \href{https://doi.org/10.1093/mnras/112.2.215}{\textcolor{violet}{\mnras}}, \href{https://ui.adsabs.harvard.edu/abs/1952MNRAS.112..215G}{112, 215}

\bibitem[{{Gordon} {et~al.}(2018){Gordon}, {Lopez-Rodriguez}, {Andersson}, {Clarke}, {Coude}, {Moullet}, {Richards}, {Shuping}, {Vacca}, \& {Yorke}}]{Gordon-etal-2018}
{Gordon}, M.~S., {Lopez-Rodriguez}, E., {Andersson}, B.~G., {et~al.} 2018, \href{https://ui.adsabs.harvard.edu/abs/2018arXiv181103100G}{{SOFIA Community Science I: HAWC+ Polarimetry of 30 Doradus}}, \href{https://arxiv.org/abs/1811.03100}{\textcolor{violet}{arXiv:1811.03100}}

\bibitem[{{Greaves} {et~al.}(2008){Greaves}, {Richards}, {Rice}, \& {Muxlow}}]{Greaves-etal-2008}
{Greaves}, J.~S., {Richards}, A.~M.~S., {Rice}, W.~K.~M., \& {Muxlow}, T.~W.~B. 2008, \href{https://doi.org/10.1111/j.1745-3933.2008.00559.x}{\textcolor{violet}{\mnras}}, \href{https://ui.adsabs.harvard.edu/abs/2008MNRAS.391L..74G}{391, L74}

\bibitem[{{Greenberg}(1968)}]{Greenberg-1968}
{Greenberg}, J.~M. 1968, in Nebulae and Interstellar Matter, ed. B.~M. {Middlehurst} \& L.~H. {Aller} (University of Chicago Press), \href{https://ui.adsabs.harvard.edu/abs/1968nim..book..221G}{221}

\bibitem[{{Guilloteau} {et~al.}(2011){Guilloteau}, {Dutrey}, {Pi{\'e}tu}, \& {Boehler}}]{Guilloteau-etal-2011}
{Guilloteau}, S., {Dutrey}, A., {Pi{\'e}tu}, V., \& {Boehler}, Y. 2011, \href{https://doi.org/10.1051/0004-6361/201015209}{\textcolor{violet}{\aap}}, \href{https://ui.adsabs.harvard.edu/abs/2011A&A...529A.105G}{529, A105}

\bibitem[{{Harper} {et~al.}(2018){Harper}, {Runyan}, {Dowell}, {Wirth}, {Amato}, {Ames}, {Amiri}, {Banks}, {Bartels}, {Benford}, {Berthoud}, {Buchanan}, {Casey}, {Chapman}, {Chuss}, {Cook}, {Derro}, {Dotson}, {Evans}, {Fixsen}, {Gatley}, {Guerra}, {Halpern}, {Hamilton}, {Hamlin}, {Hansen}, {Heimsath}, {Hermida}, {Hilton}, {Hirsch}, {Hollister}, {Hostetter}, {Irwin}, {Jhabvala}, {Jhabvala}, {Kastner}, {Kov{\'a}cs}, {Lin}, {Loewenstein}, {Looney}, {Lopez-Rodriguez}, {Maher}, {Michail}, {Miller}, {Moseley}, {Novak}, {Pernic}, {Rennick}, {Rhody}, {Sandberg}, {Sandford}, {Santos}, {Shafer}, {Sharp}, {Shirron}, {Siah}, {Silverberg}, {Sparr}, {Spotz}, {Staguhn}, {Toorian}, {Towey}, {Tuttle}, {Vaillancourt}, {Voellmer}, {Volpert}, {Wang}, \& {Wollack}}]{Harper-etal-2018}
{Harper}, D.~A., {Runyan}, M.~C., {Dowell}, C.~D., {et~al.} 2018, \href{https://doi.org/10.1142/S2251171718400081}{\textcolor{violet}{J. Astron. Instrum.}}, \href{https://ui.adsabs.harvard.edu/abs/2018JAI.....740008H}{7, 1840008}

\bibitem[{{Harris} {et~al.}(2020){Harris}, {Millman}, {van der Walt}, {Gommers}, {Virtanen}, {Cournapeau}, {Wieser}, {Taylor}, {Berg}, {Smith}, {Kern}, {Picus}, {Hoyer}, {van Kerkwijk}, {Brett}, {Haldane}, {del R{\'\i}o}, {Wiebe}, {Peterson}, {G{\'e}rard-Marchant}, {Sheppard}, {Reddy}, {Weckesser}, {Abbasi}, {Gohlke}, \& {Oliphant}}]{Harris-etal-2020}
{Harris}, C.~R., {Millman}, K.~J., {van der Walt}, S.~J., {et~al.} 2020, \href{https://doi.org/10.1038/s41586-020-2649-2}{\textcolor{violet}{\nat}}, \href{https://ui.adsabs.harvard.edu/abs/2020Natur.585..357H}{585, 357}

\bibitem[{{Harrison} {et~al.}(2021){Harrison}, {Looney}, {Stephens}, {Li}, {Teague}, {Crutcher}, {Yang}, {Cox}, {Fern{\'a}ndez-L{\'o}pez}, \& {Shinnaga}}]{Harrison-etal-2021}
{Harrison}, R.~E., {Looney}, L.~W., {Stephens}, I.~W., {et~al.} 2021, \href{https://doi.org/10.3847/1538-4357/abd94e}{\textcolor{violet}{\apj}}, \href{https://ui.adsabs.harvard.edu/abs/2021ApJ...908..141H}{908, 141}

\bibitem[{{Hartmann} {et~al.}(1998){Hartmann}, {Calvet}, {Gullbring}, \& {D'Alessio}}]{Hartmann-etal-1998}
{Hartmann}, L., {Calvet}, N., {Gullbring}, E., \& {D'Alessio}, P. 1998, \href{https://doi.org/10.1086/305277}{\textcolor{violet}{\apj}}, \href{https://ui.adsabs.harvard.edu/abs/1998ApJ...495..385H}{495, 385}

\bibitem[{{Hayashi} {et~al.}(1993){Hayashi}, {Ohashi}, \& {Miyama}}]{Hayashi-etal-1993}
{Hayashi}, M., {Ohashi}, N., \& {Miyama}, S.~M. 1993, \href{https://doi.org/10.1086/187119}{\textcolor{violet}{\apjl}}, \href{https://ui.adsabs.harvard.edu/abs/1993ApJ...418L..71H}{418, L71}

\bibitem[{{Henning} \& {Stognienko}(1996)}]{Henning-Stognienko-1996}
{Henning}, T. \& {Stognienko}, R. 1996, \aap, \href{https://ui.adsabs.harvard.edu/abs/1996A&A...311..291H}{311, 291}

\bibitem[{{Herranen} {et~al.}(2019){Herranen}, {Lazarian}, \& {Hoang}}]{Herranen-etal-2019}
{Herranen}, J., {Lazarian}, A., \& {Hoang}, T. 2019, \href{https://doi.org/10.3847/1538-4357/ab1eb3}{\textcolor{violet}{\apj}}, \href{https://ui.adsabs.harvard.edu/abs/2019ApJ...878...96H}{878, 96}

\bibitem[{{Hildebrand} \& {Dragovan}(1995)}]{Hildebrand-Dragovan-1995}
{Hildebrand}, R.~H. \& {Dragovan}, M. 1995, \href{https://doi.org/10.1086/176173}{\textcolor{violet}{\apj}}, \href{https://ui.adsabs.harvard.edu/abs/1995ApJ...450..663H}{450, 663}

\bibitem[{{Hoang} \& {Lazarian}(2009)}]{Hoang-Lazarian-2009}
{Hoang}, T. \& {Lazarian}, A. 2009, \href{https://doi.org/10.1088/0004-637X/695/2/1457}{\textcolor{violet}{\apj}}, \href{https://ui.adsabs.harvard.edu/abs/2009ApJ...695.1457H}{695, 1457}

\bibitem[{{Hoang} \& {Lazarian}(2014)}]{Hoang-Lazarian-2014}
{Hoang}, T. \& {Lazarian}, A. 2014, \href{https://doi.org/10.1093/mnras/stt2240}{\textcolor{violet}{\mnras}}, \href{https://ui.adsabs.harvard.edu/abs/2014MNRAS.438..680H}{438, 680}

\bibitem[{{Hughes} {et~al.}(2009){Hughes}, {Wilner}, {Cho}, {Marrone}, {Lazarian}, {Andrews}, \& {Rao}}]{Hughes-etal-2009}
{Hughes}, A.~M., {Wilner}, D.~J., {Cho}, J., {et~al.} 2009, \href{https://doi.org/10.1088/0004-637X/704/2/1204}{\textcolor{violet}{\apj}}, \href{https://ui.adsabs.harvard.edu/abs/2009ApJ...704.1204H}{704, 1204}

\bibitem[{{Hull} {et~al.}(2018){Hull}, {Yang}, {Li}, {Kataoka}, {Stephens}, {Andrews}, {Bai}, {Cleeves}, {Hughes}, {Looney}, {P{\'e}rez}, \& {Wilner}}]{Hull-etal-2018}
{Hull}, C. L.~H., {Yang}, H., {Li}, Z.-Y., {et~al.} 2018, \href{https://doi.org/10.3847/1538-4357/aabfeb}{\textcolor{violet}{\apj}}, \href{https://ui.adsabs.harvard.edu/abs/2018ApJ...860...82H}{860, 82}

\bibitem[{{Hull} \& {Zhang}(2019)}]{Hull-Zhang-2019}
{Hull}, C. L.~H. \& {Zhang}, Q. 2019, \href{https://doi.org/10.3389/fspas.2019.00003}{\textcolor{violet}{Front. Astron. Space Sci.}}, \href{https://ui.adsabs.harvard.edu/abs/2019FrASS...6....3H}{6, 3}

\bibitem[{{Hunter}(2007)}]{Hunter-2007}
{Hunter}, J.~D. 2007, \href{https://doi.org/10.1109/MCSE.2007.55}{\textcolor{violet}{Comput. Sci. Eng.}}, \href{https://ui.adsabs.harvard.edu/abs/2007CSE.....9...90H}{9, 90}

\bibitem[{{Kataoka} {et~al.}(2016{\natexlab{a}}){Kataoka}, {Muto}, {Momose}, {Tsukagoshi}, \& {Dullemond}}]{Kataoka-etal-2016a}
{Kataoka}, A., {Muto}, T., {Momose}, M., {Tsukagoshi}, T., \& {Dullemond}, C.~P. 2016{\natexlab{a}}, \href{https://doi.org/10.3847/0004-637X/820/1/54}{\textcolor{violet}{\apj}}, \href{https://ui.adsabs.harvard.edu/abs/2016ApJ...820...54K}{820, 54}

\bibitem[{{Kataoka} {et~al.}(2015){Kataoka}, {Muto}, {Momose}, {Tsukagoshi}, {Fukagawa}, {Shibai}, {Hanawa}, {Murakawa}, \& {Dullemond}}]{Kataoka-etal-2015}
{Kataoka}, A., {Muto}, T., {Momose}, M., {et~al.} 2015, \href{https://doi.org/10.1088/0004-637X/809/1/78}{\textcolor{violet}{\apj}}, \href{https://ui.adsabs.harvard.edu/abs/2015ApJ...809...78K}{809, 78}

\bibitem[{{Kataoka} {et~al.}(2019){Kataoka}, {Okuzumi}, \& {Tazaki}}]{Kataoka-etal-2019}
{Kataoka}, A., {Okuzumi}, S., \& {Tazaki}, R. 2019, \href{https://doi.org/10.3847/2041-8213/ab0c9a}{\textcolor{violet}{\apjl}}, \href{https://ui.adsabs.harvard.edu/abs/2019ApJ...874L...6K}{874, L6}

\bibitem[{{Kataoka} {et~al.}(2016{\natexlab{b}}){Kataoka}, {Tsukagoshi}, {Momose}, {Nagai}, {Muto}, {Dullemond}, {Pohl}, {Fukagawa}, {Shibai}, {Hanawa}, \& {Murakawa}}]{Kataoka-etal-2016b}
{Kataoka}, A., {Tsukagoshi}, T., {Momose}, M., {et~al.} 2016{\natexlab{b}}, \href{https://doi.org/10.3847/2041-8205/831/2/L12}{\textcolor{violet}{\apjl}}, \href{https://ui.adsabs.harvard.edu/abs/2016ApJ...831L..12K}{831, L12}

\bibitem[{{Kataoka} {et~al.}(2017){Kataoka}, {Tsukagoshi}, {Pohl}, {Muto}, {Nagai}, {Stephens}, {Tomisaka}, \& {Momose}}]{Kataoka-etal-2017}
{Kataoka}, A., {Tsukagoshi}, T., {Pohl}, A., {et~al.} 2017, \href{https://doi.org/10.3847/2041-8213/aa7e33}{\textcolor{violet}{\apjl}}, \href{https://ui.adsabs.harvard.edu/abs/2017ApJ...844L...5K}{844, L5}

\bibitem[{{Kenyon} {et~al.}(1994){Kenyon}, {Dobrzycka}, \& {Hartmann}}]{Kenyon-etal-1994}
{Kenyon}, S.~J., {Dobrzycka}, D., \& {Hartmann}, L. 1994, \href{https://doi.org/10.1086/117200}{\textcolor{violet}{\aj}}, \href{https://ui.adsabs.harvard.edu/abs/1994AJ....108.1872K}{108, 1872}

\bibitem[{{Kirchschlager} \& {Bertrang}(2020)}]{Kirchschlager-Bertrang-2020}
{Kirchschlager}, F. \& {Bertrang}, G. H.~M. 2020, \href{https://doi.org/10.1051/0004-6361/202037943}{\textcolor{violet}{\aap}}, \href{https://ui.adsabs.harvard.edu/abs/2020A&A...638A.116K}{638, A116}

\bibitem[{{Kirchschlager} {et~al.}(2019){Kirchschlager}, {Bertrang}, \& {Flock}}]{Kirchschlager-etal-2019}
{Kirchschlager}, F., {Bertrang}, G. H.~M., \& {Flock}, M. 2019, \href{https://doi.org/10.1093/mnras/stz1763}{\textcolor{violet}{\mnras}}, \href{https://ui.adsabs.harvard.edu/abs/2019MNRAS.488.1211K}{488, 1211}

\bibitem[{{Kirchschlager} {et~al.}(2016){Kirchschlager}, {Wolf}, \& {Madlener}}]{Kirchschlager-etal-2016}
{Kirchschlager}, F., {Wolf}, S., \& {Madlener}, D. 2016, \href{https://doi.org/10.1093/mnras/stw1692}{\textcolor{violet}{\mnras}}, \href{https://ui.adsabs.harvard.edu/abs/2016MNRAS.462..858K}{462, 858}

\bibitem[{{Kwon} {et~al.}(2011){Kwon}, {Looney}, \& {Mundy}}]{Kwon-etal-2011}
{Kwon}, W., {Looney}, L.~W., \& {Mundy}, L.~G. 2011, \href{https://doi.org/10.1088/0004-637X/741/1/3}{\textcolor{violet}{\apj}}, \href{https://ui.adsabs.harvard.edu/abs/2011ApJ...741....3K}{741, 3}

\bibitem[{{Kwon} {et~al.}(2015){Kwon}, {Looney}, {Mundy}, \& {Welch}}]{Kwon-etal-2015}
{Kwon}, W., {Looney}, L.~W., {Mundy}, L.~G., \& {Welch}, W.~J. 2015, \href{https://doi.org/10.1088/0004-637X/808/1/102}{\textcolor{violet}{\apj}}, \href{https://ui.adsabs.harvard.edu/abs/2015ApJ...808..102K}{808, 102}

\bibitem[{{Lay} {et~al.}(1997){Lay}, {Carlstrom}, \& {Hills}}]{Lay-etal-1997}
{Lay}, O.~P., {Carlstrom}, J.~E., \& {Hills}, R.~E. 1997, \href{https://doi.org/10.1086/304815}{\textcolor{violet}{\apj}}, \href{https://ui.adsabs.harvard.edu/abs/1997ApJ...489..917L}{489, 917}

\bibitem[{{Lazarian}(2007)}]{Lazarian-2007}
{Lazarian}, A. 2007, \href{https://doi.org/10.1016/j.jqsrt.2007.01.038}{\textcolor{violet}{\jqsrt}}, \href{https://ui.adsabs.harvard.edu/abs/2007JQSRT.106..225L}{106, 225}

\bibitem[{{Lazarian} \& {Hoang}(2007{\natexlab{a}})}]{Lazarian-Hoang-2007a}
{Lazarian}, A. \& {Hoang}, T. 2007{\natexlab{a}}, \href{https://doi.org/10.1111/j.1365-2966.2007.11817.x}{\textcolor{violet}{\mnras}}, \href{https://ui.adsabs.harvard.edu/abs/2007MNRAS.378..910L}{378, 910}

\bibitem[{{Lazarian} \& {Hoang}(2007{\natexlab{b}})}]{Lazarian-Hoang-2007b}
{Lazarian}, A. \& {Hoang}, T. 2007{\natexlab{b}}, \href{https://doi.org/10.1086/523849}{\textcolor{violet}{\apjl}}, \href{https://ui.adsabs.harvard.edu/abs/2007ApJ...669L..77L}{669, L77}

\bibitem[{{Lee} \& {Draine}(1985)}]{Lee-Draine-1985}
{Lee}, H.~M. \& {Draine}, B.~T. 1985, \href{https://doi.org/10.1086/162974}{\textcolor{violet}{\apj}}, \href{https://ui.adsabs.harvard.edu/abs/1985ApJ...290..211L}{290, 211}

\bibitem[{{Li} {et~al.}(2016){Li}, {Pantin}, {Telesco}, {Zhang}, {Wright}, {Barnes}, {Packham}, \& {Mari{\~n}as}}]{Li-etal-2016}
{Li}, D., {Pantin}, E., {Telesco}, C.~M., {et~al.} 2016, \href{https://doi.org/10.3847/0004-637X/832/1/18}{\textcolor{violet}{\apj}}, \href{https://ui.adsabs.harvard.edu/abs/2016ApJ...832...18L}{832, 18}

\bibitem[{{Li} {et~al.}(2018){Li}, {Telesco}, {Zhang}, {Wright}, {Pantin}, {Barnes}, \& {Packham}}]{Li-etal-2018}
{Li}, D., {Telesco}, C.~M., {Zhang}, H., {et~al.} 2018, \href{https://doi.org/10.1093/mnras/stx2228}{\textcolor{violet}{\mnras}}, \href{https://ui.adsabs.harvard.edu/abs/2018MNRAS.473.1427L}{473, 1427}

\bibitem[{{Li} {et~al.}(2014){Li}, {Banerjee}, {Pudritz}, {J{\o}rgensen}, {Shang}, {Krasnopolsky}, \& {Maury}}]{Li-etal-2014}
{Li}, Z.~Y., {Banerjee}, R., {Pudritz}, R.~E., {et~al.} 2014, in Protostars and Planets VI, ed. H.~{Beuther}, R.~S. {Klessen}, C.~P. {Dullemond}, \& T.~{Henning}, \href{https://ui.adsabs.harvard.edu/abs/2014prpl.conf..173L}{173}

\bibitem[{{Lin} {et~al.}(2024){Lin}, {Li}, {Stephens}, {Fern{\'a}ndez-L{\'o}pez}, {Carrasco-Gonz{\'a}lez}, {Chandler}, {Pasetto}, {Looney}, {Yang}, {Harrison}, {Sadavoy}, {Henning}, {Hughes}, {Kataoka}, {Kwon}, {Muto}, \& {Segura-Cox}}]{Lin-etal-2024}
{Lin}, Z.-Y.~D., {Li}, Z.-Y., {Stephens}, I.~W., {et~al.} 2024, \href{https://doi.org/10.1093/mnras/stae040}{\textcolor{violet}{\mnras}}, \href{https://ui.adsabs.harvard.edu/abs/2024MNRAS.tmp...38L}{528, 843}

\bibitem[{{Lin} {et~al.}(2023){Lin}, {Li}, {Yang}, {Mu{\~n}oz}, {Looney}, {Stephens}, {Hull}, {Fern{\'a}ndez-L{\'o}pez}, \& {Harrison}}]{Lin-etal-2023}
{Lin}, Z.-Y.~D., {Li}, Z.-Y., {Yang}, H., {et~al.} 2023, \href{https://doi.org/10.1093/mnras/stad173}{\textcolor{violet}{\mnras}}, \href{https://ui.adsabs.harvard.edu/abs/2023MNRAS.520.1210L}{520, 1210}

\bibitem[{{Liu} {et~al.}(2017){Liu}, {Henning}, {Carrasco-Gonz{\'a}lez}, {Chandler}, {Linz}, {Birnstiel}, {van Boekel}, {P{\'e}rez}, {Flock}, {Testi}, {Rodr{\'\i}guez}, \& {Galv{\'a}n-Madrid}}]{Liu-etal-2017}
{Liu}, Y., {Henning}, T., {Carrasco-Gonz{\'a}lez}, C., {et~al.} 2017, \href{https://doi.org/10.1051/0004-6361/201629786}{\textcolor{violet}{\aap}}, \href{https://ui.adsabs.harvard.edu/abs/2017A&A...607A..74L}{607, A74}

\bibitem[{{Looney} {et~al.}(2000){Looney}, {Mundy}, \& {Welch}}]{Looney-etal-2000}
{Looney}, L.~W., {Mundy}, L.~G., \& {Welch}, W.~J. 2000, \href{https://doi.org/10.1086/308239}{\textcolor{violet}{\apj}}, \href{https://ui.adsabs.harvard.edu/abs/2000ApJ...529..477L}{529, 477}

\bibitem[{{Lucas} {et~al.}(2004){Lucas}, {Fukagawa}, {Tamura}, {Beckford}, {Itoh}, {Murakawa}, {Suto}, {Hayashi}, {Oasa}, {Naoi}, {Doi}, {Ebizuka}, \& {Kaifu}}]{Lucas-etal-2004}
{Lucas}, P.~W., {Fukagawa}, M., {Tamura}, M., {et~al.} 2004, \href{https://doi.org/10.1111/j.1365-2966.2004.08026.x}{\textcolor{violet}{\mnras}}, \href{https://ui.adsabs.harvard.edu/abs/2004MNRAS.352.1347L}{352, 1347}

\bibitem[{{Lynden-Bell} \& {Pringle}(1974)}]{Lynden-Bell-Pringle-1974}
{Lynden-Bell}, D. \& {Pringle}, J.~E. 1974, \href{https://doi.org/10.1093/mnras/168.3.603}{\textcolor{violet}{\mnras}}, \href{https://ui.adsabs.harvard.edu/abs/1974MNRAS.168..603L}{168, 603}

\bibitem[{{Madlener} {et~al.}(2012){Madlener}, {Wolf}, {Dutrey}, \& {Guilloteau}}]{Madlener-etal-2012}
{Madlener}, D., {Wolf}, S., {Dutrey}, A., \& {Guilloteau}, S. 2012, \href{https://doi.org/10.1051/0004-6361/201117615}{\textcolor{violet}{\aap}}, \href{https://ui.adsabs.harvard.edu/abs/2012A&A...543A..81M}{543, A81}

\bibitem[{{Mathis} {et~al.}(1977){Mathis}, {Rumpl}, \& {Nordsieck}}]{Mathis-etal-1977}
{Mathis}, J.~S., {Rumpl}, W., \& {Nordsieck}, K.~H. 1977, \href{https://doi.org/10.1086/155591}{\textcolor{violet}{\apj}}, \href{https://ui.adsabs.harvard.edu/abs/1977ApJ...217..425M}{217, 425}

\bibitem[{{McKee} \& {Ostriker}(2007)}]{McKee-Ostriker-2007}
{McKee}, C.~F. \& {Ostriker}, E.~C. 2007, \href{https://doi.org/10.1146/annurev.astro.45.051806.110602}{\textcolor{violet}{\araa}}, \href{https://ui.adsabs.harvard.edu/abs/2007ARA&A..45..565M}{45, 565}

\bibitem[{{Men'shchikov} {et~al.}(1999){Men'shchikov}, {Henning}, \& {Fischer}}]{Menshchikov-etal-1999}
{Men'shchikov}, A.~B., {Henning}, T., \& {Fischer}, O. 1999, \href{https://doi.org/10.1086/307333}{\textcolor{violet}{\apj}}, \href{https://ui.adsabs.harvard.edu/abs/1999ApJ...519..257M}{519, 257}

\bibitem[{{Mestel}(1966)}]{Mestel-1966}
{Mestel}, L. 1966, \href{https://doi.org/10.1093/mnras/133.2.265}{\textcolor{violet}{\mnras}}, \href{https://ui.adsabs.harvard.edu/abs/1966MNRAS.133..265M}{133, 265}

\bibitem[{{Mie}(1908)}]{Mie-1908}
{Mie}, G. 1908, \href{https://doi.org/10.1002/andp.19083300302}{\textcolor{violet}{Ann. Phys.}}, \href{https://ui.adsabs.harvard.edu/abs/1908AnP...330..377M}{330, 377}

\bibitem[{{Mori} \& {Kataoka}(2021)}]{Mori-Kataoka-2021}
{Mori}, T. \& {Kataoka}, A. 2021, \href{https://doi.org/10.3847/1538-4357/abd08a}{\textcolor{violet}{\apj}}, \href{https://ui.adsabs.harvard.edu/abs/2021ApJ...908..153M}{908, 153}

\bibitem[{{Mori} {et~al.}(2019){Mori}, {Kataoka}, {Ohashi}, {Momose}, {Muto}, {Nagai}, \& {Tsukagoshi}}]{Mori-etal-2019}
{Mori}, T., {Kataoka}, A., {Ohashi}, S., {et~al.} 2019, \href{https://doi.org/10.3847/1538-4357/ab3575}{\textcolor{violet}{\apj}}, \href{https://ui.adsabs.harvard.edu/abs/2019ApJ...883...16M}{883, 16}

\bibitem[{{Mullin} {et~al.}(2024){Mullin}, {Dong}, {Leisenring}, {Cugno}, {Greene}, {Johnstone}, {Meyer}, {Wagner}, {Wolff}, {Boyer}, {Horner}, {Hodapp}, {McCarthy}, {Rieke}, {Rieke}, \& {Young}}]{Mullin-etal-2024}
{Mullin}, C., {Dong}, R., {Leisenring}, J., {et~al.} 2024, \href{https://doi.org/10.3847/1538-3881/ad2de9}{\textcolor{violet}{\aj}}, \href{https://ui.adsabs.harvard.edu/abs/2024AJ....167..183M}{167, 183}

\bibitem[{{Mundt} {et~al.}(1990){Mundt}, {Buehrke}, {Solf}, {Ray}, \& {Raga}}]{Mundt-etal-1990}
{Mundt}, R., {Buehrke}, T., {Solf}, J., {Ray}, T.~P., \& {Raga}, A.~C. 1990, \aap, \href{https://ui.adsabs.harvard.edu/abs/1990A&A...232...37M}{232, 37}

\bibitem[{{Mundt} \& {Fried}(1983)}]{Mundt-Fried-1983}
{Mundt}, R. \& {Fried}, J.~W. 1983, \href{https://doi.org/10.1086/184155}{\textcolor{violet}{\apjl}}, \href{https://ui.adsabs.harvard.edu/abs/1983ApJ...274L..83M}{274, L83}

\bibitem[{{Mundy} {et~al.}(1996){Mundy}, {Looney}, {Erickson}, {Grossman}, {Welch}, {Forster}, {Wright}, {Plambeck}, {Lugten}, \& {Thornton}}]{Mundy-etal-1996}
{Mundy}, L.~G., {Looney}, L.~W., {Erickson}, W., {et~al.} 1996, \href{https://doi.org/10.1086/310117}{\textcolor{violet}{\apjl}}, \href{https://ui.adsabs.harvard.edu/abs/1996ApJ...464L.169M}{464, L169}

\bibitem[{{Murakawa} {et~al.}(2008){Murakawa}, {Oya}, {Pyo}, \& {Ishii}}]{Murakawa-etal-2008}
{Murakawa}, K., {Oya}, S., {Pyo}, T.~S., \& {Ishii}, M. 2008, \href{https://doi.org/10.1051/0004-6361:200810723}{\textcolor{violet}{\aap}}, \href{https://ui.adsabs.harvard.edu/abs/2008A&A...492..731M}{492, 731}

\bibitem[{{Ohashi} \& {Kataoka}(2019)}]{Ohashi-Kataoka-2019}
{Ohashi}, S. \& {Kataoka}, A. 2019, \href{https://doi.org/10.3847/1538-4357/ab5107}{\textcolor{violet}{\apj}}, \href{https://ui.adsabs.harvard.edu/abs/2019ApJ...886..103O}{886, 103}

\bibitem[{{Ohashi} {et~al.}(2018){Ohashi}, {Kataoka}, {Nagai}, {Momose}, {Muto}, {Hanawa}, {Fukagawa}, {Tsukagoshi}, {Murakawa}, \& {Shibai}}]{Ohashi-etal-2018}
{Ohashi}, S., {Kataoka}, A., {Nagai}, H., {et~al.} 2018, \href{https://doi.org/10.3847/1538-4357/aad632}{\textcolor{violet}{\apj}}, \href{https://ui.adsabs.harvard.edu/abs/2018ApJ...864...81O}{864, 81}

\bibitem[{{Pattle} {et~al.}(2023){Pattle}, {Fissel}, {Tahani}, {Liu}, \& {Ntormousi}}]{Pattle-etal-2023}
{Pattle}, K., {Fissel}, L., {Tahani}, M., {Liu}, T., \& {Ntormousi}, E. 2023, in Astronomical Society of the Pacific Conference Series, Vol. 534, Protostars and Planets VII, ed. S.~{Inutsuka}, Y.~{Aikawa}, T.~{Muto}, K.~{Tomida}, \& M.~{Tamura}, \href{https://ui.adsabs.harvard.edu/abs/2023ASPC..534..193P}{193}

\bibitem[{{Pinte} {et~al.}(2016){Pinte}, {Dent}, {M{\'e}nard}, {Hales}, {Hill}, {Cortes}, \& {de Gregorio-Monsalvo}}]{Pinte-etal-2016}
{Pinte}, C., {Dent}, W.~R.~F., {M{\'e}nard}, F., {et~al.} 2016, \href{https://doi.org/10.3847/0004-637X/816/1/25}{\textcolor{violet}{\apj}}, \href{https://ui.adsabs.harvard.edu/abs/2016ApJ...816...25P}{816, 25}

\bibitem[{{Pollack} {et~al.}(1994){Pollack}, {Hollenbach}, {Beckwith}, {Simonelli}, {Roush}, \& {Fong}}]{Pollack-etal-1994}
{Pollack}, J.~B., {Hollenbach}, D., {Beckwith}, S., {et~al.} 1994, \href{https://doi.org/10.1086/173677}{\textcolor{violet}{\apj}}, \href{https://ui.adsabs.harvard.edu/abs/1994ApJ...421..615P}{421, 615}

\bibitem[{{Pringle}(1981)}]{Pringle-1981}
{Pringle}, J.~E. 1981, \href{https://doi.org/10.1146/annurev.aa.19.090181.001033}{\textcolor{violet}{\araa}}, \href{https://ui.adsabs.harvard.edu/abs/1981ARA&A..19..137P}{19, 137}

\bibitem[{{Rebull} {et~al.}(2004){Rebull}, {Wolff}, \& {Strom}}]{Rebull-etal-2004}
{Rebull}, L.~M., {Wolff}, S.~C., \& {Strom}, S.~E. 2004, \href{https://doi.org/10.1086/380931}{\textcolor{violet}{\aj}}, \href{https://ui.adsabs.harvard.edu/abs/2004AJ....127.1029R}{127, 1029}

\bibitem[{{Reissl} {et~al.}(2020){Reissl}, {Guillet}, {Brauer}, {Levrier}, {Boulanger}, \& {Klessen}}]{Reissl-etal-2020}
{Reissl}, S., {Guillet}, V., {Brauer}, R., {et~al.} 2020, \href{https://doi.org/10.1051/0004-6361/201937177}{\textcolor{violet}{\aap}}, \href{https://ui.adsabs.harvard.edu/abs/2020A&A...640A.118R}{640, A118}

\bibitem[{{Reissl} {et~al.}(2017){Reissl}, {Seifried}, {Wolf}, {Banerjee}, \& {Klessen}}]{Reissl-etal-2017}
{Reissl}, S., {Seifried}, D., {Wolf}, S., {Banerjee}, R., \& {Klessen}, R.~S. 2017, \href{https://doi.org/10.1051/0004-6361/201730408}{\textcolor{violet}{\aap}}, \href{https://ui.adsabs.harvard.edu/abs/2017A&A...603A..71R}{603, A71}

\bibitem[{{Reissl} {et~al.}(2016){Reissl}, {Wolf}, \& {Brauer}}]{Reissl-etal-2016}
{Reissl}, S., {Wolf}, S., \& {Brauer}, R. 2016, \href{https://doi.org/10.1051/0004-6361/201424930}{\textcolor{violet}{\aap}}, \href{https://ui.adsabs.harvard.edu/abs/2016A&A...593A..87R}{593, A87}

\bibitem[{{Reissl} {et~al.}(2018){Reissl}, {Wolf}, \& {Brauer}}]{Reissl-etal-2018}
{Reissl}, S., {Wolf}, S., \& {Brauer}, R. 2018, \href{https://ui.adsabs.harvard.edu/abs/2018ascl.soft07001R}{{POLARIS: POLArized RadIation Simulator}}, \href{https://ascl.net/1807.001}{\textcolor{violet}{ascl:1807.001}}

\bibitem[{{Roberge} \& {Lazarian}(1999)}]{Roberge-Lazarian-1999}
{Roberge}, W.~G. \& {Lazarian}, A. 1999, \href{https://doi.org/10.1046/j.1365-8711.1999.02464.x}{\textcolor{violet}{\mnras}}, \href{https://ui.adsabs.harvard.edu/abs/1999MNRAS.305..615R}{305, 615}

\bibitem[{{Robitaille} \& {Bressert}(2012)}]{Robitaille-Bressert-2012}
{Robitaille}, T. \& {Bressert}, E. 2012, \href{https://ui.adsabs.harvard.edu/abs/2012ascl.soft08017R}{{APLpy: Astronomical Plotting Library in Python}}, \href{https://ascl.net/1208.017}{\textcolor{violet}{ascl:1208.017}}

\bibitem[{{Robitaille} {et~al.}(2007){Robitaille}, {Whitney}, {Indebetouw}, \& {Wood}}]{Robitaille-etal-2007}
{Robitaille}, T.~P., {Whitney}, B.~A., {Indebetouw}, R., \& {Wood}, K. 2007, \href{https://doi.org/10.1086/512039}{\textcolor{violet}{\apjs}}, \href{https://ui.adsabs.harvard.edu/abs/2007ApJS..169..328R}{169, 328}

\bibitem[{{Seifried} {et~al.}(2020){Seifried}, {Walch}, {Weis}, {Reissl}, {Soler}, {Klessen}, \& {Joshi}}]{Seifried-etal-2020}
{Seifried}, D., {Walch}, S., {Weis}, M., {et~al.} 2020, \href{https://doi.org/10.1093/mnras/staa2231}{\textcolor{violet}{\mnras}}, \href{https://ui.adsabs.harvard.edu/abs/2020MNRAS.497.4196S}{497, 4196}

\bibitem[{{Stapelfeldt} {et~al.}(1995){Stapelfeldt}, {Burrows}, {Krist}, {Trauger}, {Hester}, {Holtzman}, {Ballester}, {Casertano}, {Clarke}, {Crisp}, {Evans}, {Gallagher}, {Griffiths}, {Hoessel}, {Mould}, {Scowen}, {Watson}, \& {Westphal}}]{Stapelfeldt-etal-1995}
{Stapelfeldt}, K.~R., {Burrows}, C.~J., {Krist}, J.~E., {et~al.} 1995, \href{https://doi.org/10.1086/176106}{\textcolor{violet}{\apj}}, \href{https://ui.adsabs.harvard.edu/abs/1995ApJ...449..888S}{449, 888}

\bibitem[{{Stephens} {et~al.}(2023){Stephens}, {Lin}, {Fern{\'a}ndez-L{\'o}pez}, {Li}, {Looney}, {Yang}, {Harrison}, {Kataoka}, {Carrasco-Gonzalez}, {Okuzumi}, \& {Tazaki}}]{Stephens-etal-2023}
{Stephens}, I.~W., {Lin}, Z.-Y.~D., {Fern{\'a}ndez-L{\'o}pez}, M., {et~al.} 2023, \href{https://doi.org/10.1038/s41586-023-06648-7}{\textcolor{violet}{\nat}}, \href{https://ui.adsabs.harvard.edu/abs/2023Natur.623..705S}{623, 705}

\bibitem[{{Stephens} {et~al.}(2014){Stephens}, {Looney}, {Kwon}, {Fern{\'a}ndez-L{\'o}pez}, {Hughes}, {Mundy}, {Crutcher}, {Li}, \& {Rao}}]{Stephens-etal-2014}
{Stephens}, I.~W., {Looney}, L.~W., {Kwon}, W., {et~al.} 2014, \href{https://doi.org/10.1038/nature13850}{\textcolor{violet}{\nat}}, \href{https://ui.adsabs.harvard.edu/abs/2014Natur.514..597S}{514, 597}

\bibitem[{{Stephens} {et~al.}(2017){Stephens}, {Yang}, {Li}, {Looney}, {Kataoka}, {Kwon}, {Fern{\'a}ndez-L{\'o}pez}, {Hull}, {Hughes}, {Segura-Cox}, {Mundy}, {Crutcher}, \& {Rao}}]{Stephens-etal-2017}
{Stephens}, I.~W., {Yang}, H., {Li}, Z.-Y., {et~al.} 2017, \href{https://doi.org/10.3847/1538-4357/aa998b}{\textcolor{violet}{\apj}}, \href{https://ui.adsabs.harvard.edu/abs/2017ApJ...851...55S}{851, 55}

\bibitem[{{Tazaki} {et~al.}(2017){Tazaki}, {Lazarian}, \& {Nomura}}]{Tazaki-etal-2017}
{Tazaki}, R., {Lazarian}, A., \& {Nomura}, H. 2017, \href{https://doi.org/10.3847/1538-4357/839/1/56}{\textcolor{violet}{\apj}}, \href{https://ui.adsabs.harvard.edu/abs/2017ApJ...839...56T}{839, 56}

\bibitem[{{Tazaki} {et~al.}(2019{\natexlab{a}}){Tazaki}, {Tanaka}, {Kataoka}, {Okuzumi}, \& {Muto}}]{Tazaki-etal-2019a}
{Tazaki}, R., {Tanaka}, H., {Kataoka}, A., {Okuzumi}, S., \& {Muto}, T. 2019{\natexlab{a}}, \href{https://doi.org/10.3847/1538-4357/ab45f0}{\textcolor{violet}{\apj}}, \href{https://ui.adsabs.harvard.edu/abs/2019ApJ...885...52T}{885, 52}

\bibitem[{{Tazaki} {et~al.}(2019{\natexlab{b}}){Tazaki}, {Tanaka}, {Muto}, {Kataoka}, \& {Okuzumi}}]{Tazaki-etal-2019b}
{Tazaki}, R., {Tanaka}, H., {Muto}, T., {Kataoka}, A., \& {Okuzumi}, S. 2019{\natexlab{b}}, \href{https://doi.org/10.1093/mnras/stz662}{\textcolor{violet}{\mnras}}, \href{https://ui.adsabs.harvard.edu/abs/2019MNRAS.485.4951T}{485, 4951}

\bibitem[{{Temi} {et~al.}(2018){Temi}, {Hoffman}, {Ennico}, \& {Le}}]{Temi-etal-2018}
{Temi}, P., {Hoffman}, D., {Ennico}, K., \& {Le}, J. 2018, \href{https://doi.org/10.1142/S2251171718400111}{\textcolor{violet}{J. Astron. Instrum.}}, \href{https://ui.adsabs.harvard.edu/abs/2018JAI.....740011T}{7, 1840011}

\bibitem[{{Tsukamoto}(2016)}]{Tsukamoto-2016}
{Tsukamoto}, Y. 2016, \href{https://doi.org/10.1017/pasa.2016.6}{\textcolor{violet}{\pasa}}, \href{https://ui.adsabs.harvard.edu/abs/2016PASA...33...10T}{33, e010}

\bibitem[{{Ueda} {et~al.}(2021){Ueda}, {Kataoka}, {Zhang}, {Zhu}, {Carrasco-Gonz{\'a}lez}, \& {Sierra}}]{Ueda-etal-2021}
{Ueda}, T., {Kataoka}, A., {Zhang}, S., {et~al.} 2021, \href{https://doi.org/10.3847/1538-4357/abf7b8}{\textcolor{violet}{\apj}}, \href{https://ui.adsabs.harvard.edu/abs/2021ApJ...913..117U}{913, 117}

\bibitem[{{Wardle} \& {Kronberg}(1974)}]{Wardle-Kronberg-1974}
{Wardle}, J.~F.~C. \& {Kronberg}, P.~P. 1974, \href{https://doi.org/10.1086/153240}{\textcolor{violet}{\apj}}, \href{https://ui.adsabs.harvard.edu/abs/1974ApJ...194..249W}{194, 249}

\bibitem[{{Warren} \& {Brandt}(2008)}]{Warren-Brandt-2008}
{Warren}, S.~G. \& {Brandt}, R.~E. 2008, \href{https://doi.org/10.1029/2007JD009744}{\textcolor{violet}{J. Geophys. Res.}}, \href{https://ui.adsabs.harvard.edu/abs/2008JGRD..11314220W}{113, D14220}

\bibitem[{{Wolf} {et~al.}(2003){Wolf}, {Padgett}, \& {Stapelfeldt}}]{Wolf-etal-2003}
{Wolf}, S., {Padgett}, D.~L., \& {Stapelfeldt}, K.~R. 2003, \href{https://doi.org/10.1086/374041}{\textcolor{violet}{\apj}}, \href{https://ui.adsabs.harvard.edu/abs/2003ApJ...588..373W}{588, 373}

\bibitem[{{Wolf} \& {Voshchinnikov}(2004)}]{Wolf-Voshchinnikov-2004}
{Wolf}, S. \& {Voshchinnikov}, N.~V. 2004, \href{https://doi.org/10.1016/j.cpc.2004.06.070}{\textcolor{violet}{Comput. Phys. Commun.}}, \href{https://ui.adsabs.harvard.edu/abs/2004CoPhC.162..113W}{162, 113}

\bibitem[{{Wolf} \& {Voshchinnikov}(2018)}]{Wolf-Voshchinnikov-2018}
{Wolf}, S. \& {Voshchinnikov}, N.~V. 2018, \href{https://ui.adsabs.harvard.edu/abs/2018ascl.soft10019W}{{MIEX: Mie scattering code for large grains}}, \href{https://ascl.net/1810.019}{\textcolor{violet}{ascl:1810.019}}

\bibitem[{{Wurster} \& {Li}(2018)}]{Wurster-Li-2018}
{Wurster}, J. \& {Li}, Z.-Y. 2018, \href{https://doi.org/10.3389/fspas.2018.00039}{\textcolor{violet}{Front. Astron. Space Sci.}}, \href{https://ui.adsabs.harvard.edu/abs/2018FrASS...5...39W}{5, 39}

\bibitem[{{Yang} {et~al.}(2016){Yang}, {Li}, {Looney}, \& {Stephens}}]{Yang-etal-2016}
{Yang}, H., {Li}, Z.-Y., {Looney}, L., \& {Stephens}, I. 2016, \href{https://doi.org/10.1093/mnras/stv2633}{\textcolor{violet}{\mnras}}, \href{https://ui.adsabs.harvard.edu/abs/2016MNRAS.456.2794Y}{456, 2794}

\bibitem[{{Yang} {et~al.}(2019){Yang}, {Li}, {Stephens}, {Kataoka}, \& {Looney}}]{Yang-etal-2019}
{Yang}, H., {Li}, Z.-Y., {Stephens}, I.~W., {Kataoka}, A., \& {Looney}, L. 2019, \href{https://doi.org/10.1093/mnras/sty3263}{\textcolor{violet}{\mnras}}, \href{https://ui.adsabs.harvard.edu/abs/2019MNRAS.483.2371Y}{483, 2371}

\bibitem[{{Zielinski} \& {Wolf}(2022)}]{Zielinski-Wolf-2022}
{Zielinski}, N. \& {Wolf}, S. 2022, \href{https://doi.org/10.1051/0004-6361/202141537}{\textcolor{violet}{\aap}}, \href{https://ui.adsabs.harvard.edu/abs/2022A&A...659A..22Z}{659, A22}

\bibitem[{{Zielinski} {et~al.}(2021){Zielinski}, {Wolf}, \& {Brunngr{\"a}ber}}]{Zielinski-etal-2021}
{Zielinski}, N., {Wolf}, S., \& {Brunngr{\"a}ber}, R. 2021, \href{https://doi.org/10.1051/0004-6361/202039126}{\textcolor{violet}{\aap}}, \href{https://ui.adsabs.harvard.edu/abs/2021A&A...645A.125Z}{645, A125}

\end{thebibliography}
% \listofobjects

\end{document}